\batchmode
\makeatletter
\def\input@path{{"/home/jacob/Documents/Work/My Papers/2026-Against the Ontological Wave Function in Bohmian Mechanics/"}}
\makeatother
\documentclass[11pt,twoside,english]{article}
\usepackage[T1]{fontenc}
\usepackage[latin9]{inputenc}
\usepackage{babel}
\usepackage{amsmath}
\usepackage{amssymb}
\usepackage{cancel}
\usepackage{geometry}
\usepackage{setspace}
\usepackage[pdftex,pdfusetitle,
 bookmarks=true,bookmarksnumbered=true,bookmarksopen=false,
 breaklinks=false,pdfborder={0 0 0},pdfborderstyle={},backref=false,colorlinks=false]
 {hyperref}

\makeatletter

\let\originalleft\left
\let\originalright\right
\renewcommand{\left}{\mathopen{}\mathclose\bgroup\originalleft}
\renewcommand{\right}{\aftergroup\egroup\originalright}

\def\smalloverbrace#1{\mathop{\vbox{\m@th\ialign{##\crcr%
      \noalign{\kern3\p@}%
      \tiny\downbracefill\crcr\noalign{\kern3\p@\nointerlineskip}%
      $\hfil\displaystyle{#1}\hfil$\crcr}}}\limits}

\def\smallunderbrace#1{\mathop{\vtop{\m@th\ialign{##\crcr
   $\hfil\displaystyle{#1}\hfil$\crcr
   \noalign{\kern3\p@\nointerlineskip}%
   \tiny\upbracefill\crcr\noalign{\kern3\p@}}}}\limits}

\DeclareMathAlphabet{\mymathbb}{U}{bbold}{m}{n}

\makeatother

\begin{document}
\title{Pilot-Wave Theories as Hidden Markov Models}
\author{Jacob A. Barandes\thanks{Departments of Philosophy and Physics, Harvard University, Cambridge, MA 02138; jacob\_barandes@harvard.edu; ORCID: 0000-0002-3740-4418}
}
\date{\today}

\maketitle

\begin{abstract}
The original version of the de Broglie-Bohm pilot-wave theory, also
called Bohmian mechanics, attempted to treat the wave function or
pilot wave as a part of the physical ontology of nature. More recent
versions of the de Broglie-Bohm theory appearing in the last few decades
have tried to regard the pilot wave instead as an aspect of the theory's
nomology, or dynamical laws. This paper argues that neither of these
views is correct, and that the de Broglie-Bohm pilot wave is best
understood as a collection of latent variables in the sense of a hidden
Markov model, a construct that was not available when de Broglie and
Bohm originally formulated what became their pilot-wave theory. This
paper also discusses several other challenges for the ontological
view of the pilot wave. One such challenge is due to Foldy-Wouthuysen
gauge transformations, which connect up with the Deotto-Ghirardi ambiguity
in the de Broglie-Bohm theory. Another challenge arises from the freedom
to carry out canonical transformations in the wave function's own
notion of phase space, as defined by Strocchi and Heslot.
\end{abstract}

\begin{center}
\global\long\def\quote#1{``#1"}%
\global\long\def\apostrophe{\textrm{'}}%
\global\long\def\slot{\phantom{x}}%
\global\long\def\eval#1{\left.#1\right\vert }%
\global\long\def\keyeq#1{\boxed{#1}}%
\global\long\def\importanteq#1{\boxed{\boxed{#1}}}%
\global\long\def\given{\vert}%
\global\long\def\mapping#1#2#3{#1:#2\to#3}%
\global\long\def\composition{\circ}%
\global\long\def\set#1{\left\{  #1\right\}  }%
\global\long\def\setindexed#1#2{\left\{  #1\right\}  _{#2}}%

\global\long\def\setbuild#1#2{\left\{  \left.\!#1\,\right|\,#2\right\}  }%
\global\long\def\complem{\mathrm{c}}%

\global\long\def\union{\cup}%
\global\long\def\intersection{\cap}%
\global\long\def\cartesianprod{\times}%
\global\long\def\disjointunion{\sqcup}%

\global\long\def\isomorphic{\cong}%

\global\long\def\setsize#1{\left|#1\right|}%
\global\long\def\defeq{\equiv}%
\global\long\def\conj{\ast}%
\global\long\def\overconj#1{\overline{#1}}%
\global\long\def\re{\mathrm{Re\,}}%
\global\long\def\im{\mathrm{Im\,}}%

\global\long\def\transp{\mathrm{T}}%
\global\long\def\tr{\mathrm{tr}}%
\global\long\def\adj{\dagger}%
\global\long\def\diag#1{\mathrm{diag}\left(#1\right)}%
\global\long\def\dotprod{\cdot}%
\global\long\def\crossprod{\times}%
\global\long\def\Probability#1{\mathrm{Prob}\left(#1\right)}%
\global\long\def\Amplitude#1{\mathrm{Amp}\left(#1\right)}%
\global\long\def\cov{\mathrm{cov}}%
\global\long\def\corr{\mathrm{corr}}%

\global\long\def\absval#1{\left\vert #1\right\vert }%
\global\long\def\expectval#1{\left\langle #1\right\rangle }%
\global\long\def\op#1{\hat{#1}}%

\global\long\def\bra#1{\left\langle #1\right|}%
\global\long\def\ket#1{\left|#1\right\rangle }%
\global\long\def\braket#1#2{\left\langle \left.\!#1\right|#2\right\rangle }%

\global\long\def\parens#1{(#1)}%
\global\long\def\bigparens#1{\big(#1\big)}%
\global\long\def\Bigparens#1{\Big(#1\Big)}%
\global\long\def\biggparens#1{\bigg(#1\bigg)}%
\global\long\def\Biggparens#1{\Bigg(#1\Bigg)}%
\global\long\def\bracks#1{[#1]}%
\global\long\def\bigbracks#1{\big[#1\big]}%
\global\long\def\Bigbracks#1{\Big[#1\Big]}%
\global\long\def\biggbracks#1{\bigg[#1\bigg]}%
\global\long\def\Biggbracks#1{\Bigg[#1\Bigg]}%
\global\long\def\curlies#1{\{#1\}}%
\global\long\def\bigcurlies#1{\big\{#1\big\}}%
\global\long\def\Bigcurlies#1{\Big\{#1\Big\}}%
\global\long\def\biggcurlies#1{\bigg\{#1\bigg\}}%
\global\long\def\Biggcurlies#1{\Bigg\{#1\Bigg\}}%
\global\long\def\verts#1{\vert#1\vert}%
\global\long\def\bigverts#1{\big\vert#1\big\vert}%
\global\long\def\Bigverts#1{\Big\vert#1\Big\vert}%
\global\long\def\biggverts#1{\bigg\vert#1\bigg\vert}%
\global\long\def\Biggverts#1{\Bigg\vert#1\Bigg\vert}%
\global\long\def\Verts#1{\Vert#1\Vert}%
\global\long\def\bigVerts#1{\big\Vert#1\big\Vert}%
\global\long\def\BigVerts#1{\Big\Vert#1\Big\Vert}%
\global\long\def\biggVerts#1{\bigg\Vert#1\bigg\Vert}%
\global\long\def\BiggVerts#1{\Bigg\Vert#1\Bigg\Vert}%
\global\long\def\ket#1{\vert#1\rangle}%
\global\long\def\bigket#1{\big\vert#1\big\rangle}%
\global\long\def\Bigket#1{\Big\vert#1\Big\rangle}%
\global\long\def\biggket#1{\bigg\vert#1\bigg\rangle}%
\global\long\def\Biggket#1{\Bigg\vert#1\Bigg\rangle}%
\global\long\def\bra#1{\langle#1\vert}%
\global\long\def\bigbra#1{\big\langle#1\big\vert}%
\global\long\def\Bigbra#1{\Big\langle#1\Big\vert}%
\global\long\def\biggbra#1{\bigg\langle#1\bigg\vert}%
\global\long\def\Biggbra#1{\Bigg\langle#1\Bigg\vert}%
\global\long\def\braket#1#2{\langle#1\vert#2\rangle}%
\global\long\def\bigbraket#1#2{\big\langle#1\big\vert#2\big\rangle}%
\global\long\def\Bigbraket#1#2{\Big\langle#1\Big\vert#2\Big\rangle}%
\global\long\def\biggbraket#1#2{\bigg\langle#1\bigg\vert#2\bigg\rangle}%
\global\long\def\Biggbraket#1#2{\Bigg\langle#1\Bigg\vert#2\Bigg\rangle}%
\global\long\def\angs#1{\langle#1\rangle}%
\global\long\def\bigangs#1{\big\langle#1\big\rangle}%
\global\long\def\Bigangs#1{\Big\langle#1\Big\rangle}%
\global\long\def\biggangs#1{\bigg\langle#1\bigg\rangle}%
\global\long\def\Biggangs#1{\Bigg\langle#1\Bigg\rangle}%

\global\long\def\vec#1{\mathbf{#1}}%
\global\long\def\vecgreek#1{\boldsymbol{#1}}%
\global\long\def\idmatrix{\mymathbb{1}}%
\global\long\def\projector{P}%
\global\long\def\permutationmatrix{\Sigma}%
\global\long\def\densitymatrix{\rho}%
\global\long\def\krausmatrix{K}%
\global\long\def\stochasticmatrix{\Gamma}%
\global\long\def\lindbladmatrix{L}%
\global\long\def\dynop{\Theta}%
\global\long\def\timeevop{U}%
\global\long\def\hadamardprod{\odot}%
\global\long\def\tensorprod{\otimes}%

\global\long\def\inprod#1#2{\left\langle #1,#2\right\rangle }%
\global\long\def\normket#1{\left\Vert #1\right\Vert }%
\global\long\def\hilbspace{\mathcal{H}}%
\global\long\def\samplespace{\Omega}%
\global\long\def\configspace{\mathcal{C}}%
\global\long\def\phasespace{\mathcal{P}}%
\global\long\def\spectrum{\sigma}%
\global\long\def\restrict#1#2{\left.#1\right\vert _{#2}}%
\global\long\def\from{\leftarrow}%
\global\long\def\statemap{\omega}%
\global\long\def\degangle#1{#1^{\circ}}%
\global\long\def\trivialvector{\tilde{v}}%
\global\long\def\eqsbrace#1{\left.#1\qquad\right\}  }%
\global\long\def\operator#1{\operatorname{#1}}%
\par\end{center}

\section{Introduction\label{sec:Introduction}}

\subsection{The Interpretation of the Quantum State\label{subsec:The-Interpretation-of-the-Quantum-State}}

The quantum state is often taken to be the starring protagonist of
quantum theory. The earliest version of the quantum state, introduced
by Erwin Schrödinger in 1926, took the form of a wave function defined
in the abstract space of a system's kinematically possible configurations,
and evolved in time according to a specific partial differential equation\textemdash the
famous Schrödinger equation (Schrödinger 1926a\textendash d)\nocite{Schrodinger:1926qae1,Schrodinger:1926qae2,Schrodinger:1926qae3,Schrodinger:1926qae4}.
Later versions of the quantum state, introduced by Paul Dirac and
John von Neumann, were represented either by state vectors in abstract
Hilbert spaces, or by statistical operators or density operators on
Hilbert spaces (Dirac 1930, von Neumann 1932)\nocite{Dirac:1930pofm,vonNeumann:1932mgdq}.
By the 1940s, quantum states had been abstracted further to positive,
normalized linear functionals in the dual spaces to C{*}-algebras
(Segal 1947a,b)\nocite{Segal:1947otionritroooahs,Segal:1947pfgqm}.

Understandably, much of the discourse surrounding the interpretation
of quantum theory over the past century has centered on the meaning
and metaphysical status of the quantum state. Contemporary postures
toward the quantum state typically  lean toward one of the following
views: 
\begin{enumerate}
\item (Statistical) The quantum state is an instrumentalist statement about
statistical distributions of possible measurement outcomes, as on
the orthodox or \textquoteleft textbook\textquoteright{} interpretation
of quantum theory. (See, for example, Shankar 1994; Sakurai, Napolitano
2010; Griffiths, Schroeter 2018.)\nocite{Shankar:1994pqm,SakuraiNapolitano:2010mqm,GriffithsSchroeter:2018iqm}
\item (Epistemic-over-measurements) The quantum state is a representation
of the epistemology or knowledge of external observers about possible
measurement results, as on the Copenhagen interpretation (Heisenberg
1955, 1958; Howard 2004)\nocite{Heisenberg:1955tdotiotqt,Heisenberg:1958paptrims,Howard:2004witciasim}
and on QBism (Fuchs 2010)\nocite{Fuchs:2010qbpqb}.
\item (Epistemic-over-ontology) The quantum state is a representation of
the epistemology or knowledge of external observers about objective
arrangements of physical or ontological entities, as on various ``$\psi$-epistemic''
accounts (Harrigan, Spekkens 2010)\nocite{HarriganSpekkens:2010eievqs}.
\item (Ontological-monistic) The quantum state is the sole ontology of the
universe, as on the Everett \textquoteleft many worlds\textquoteright{}
interpretation (Everett 1956, 1957a; Everett, DeWitt, Graham 1973)\nocite{Everett:1956ttotuwf,Everett:1957rsfqm,EverettDeWittGraham:1973mwiqm}
and on certain interpretations of dynamical-collapse theories (Gisin
1984; Ghirardi, Rimini, Weber 1986)\nocite{Gisin:1984qmasp,GhirardiRiminiWeber:1986udmms}.\footnote{At least according to Bell's account of the original GRW theory (Bell
1987)\nocite{Bell:1987atqj}, the theory did not propose any ontology
above and beyond the wave function. The ``quantum jumps'' of the
GRW theory were, in Bell's words, ``part of the wavefunction, not
something else.'' In the same paper, Bell also wrote that ``The
GRW theory does not add variables.'' In later work, Bell wrote that
``The GRW-type theories have nothing in their kinematics but the
wavefunction'' (Bell 1990)\nocite{Bell:1990am}.}
\item (Ontological-pluralistic) The quantum state is a part of the physical
reality or ontology of the universe, as on some versions of the de
Broglie-Bohm pilot-wave theory (de Broglie 1930; Bohm 1952a,b)\nocite{DeBroglie:1930ialedlmo,Bohm:1952siqtthvi,Bohm:1952siqtthvii}
and on more recent formulations of dynamical-collapse theories (Allori,
Goldstein, Tumolka, Zanghì 2008; Goldstein, Tumulka, Zanghì 2012)\nocite{AlloriGoldsteinTumulkaZanghi:2008otcsobmatgrwt,GoldsteinTumulkaZanghi:2012tqfatgf}.
\item (Nomological) The quantum state is nomological, meaning a feature
of a system's dynamical laws, as on other versions of the de Broglie-Bohm
pilot-wave theory (Dürr, Goldstein, Zanghì 1996)\nocite{DurrGoldsteinZanghi:1996bmatmotwf}.
\end{enumerate}

Arguably none of these views capture the full and proper meaning of
the quantum state. All of them seem at least somewhat ill-fitting. 

The first two views in the list above\textemdash the statistical and
epistemic-over-measurement views\textemdash leave the measurement
problem essentially unanswered (Maudlin 1995)\nocite{Maudlin:1995tmp},
and remain quiet over whether there is any ontology in the world
beyond macroscopic measuring devices and external observers. The epistemic-over-ontology
view does not seem to capture the functional relationship that quantum
states have with the behavior of systems and the outcomes of measurements,
and runs into difficulties in accommodating various no-go theorems,
like the PBR theorem (Pusey, Barrett, Rudolph 2012)\nocite{PuseyBarrettRudolph:2012rqs}.
As for the fourth and fifth ontological views, quantum states are
associated with abstract spaces, like configuration spaces or Hilbert
spaces, and have many other exotic properties that seem quite different
from the sorts of entities traditionally assigned a physical or ontological
meaning, as this paper will explain in detail, and as explored from
a historical perspective in other work (Barandes 2026b)\nocite{Barandes:2026hdotprotwf}.
Finally, looking at the nomological view, quantum states have seemingly
contingent, possibly complicated initial conditions of their own,
and also typically feature highly complicated forms of time-dependence,
as well as nontrivial behavior under time-reversal transformations,
all of which make them awkward as dynamical laws. 

\subsection{Markovian Theories and Hidden Markov Models\label{subsec:Markovian-Theories-and-Hidden-Markov-Models}}

With sufficient effort, one can mitigate most of the problems listed
above to varying degrees, but the very need for that effort also motivates
seeking out a more natural interpretation of quantum states. To that
end, this paper will argue that quantum states are best understood
through the lens of hidden Markov models, with quantum states playing
the role of their latent variables\textemdash sometimes called \textquoteleft hidden
variables,\textquoteright{} although that term should not be confused
with the conventional notion of hidden variables in quantum theory. 

In brief, for a physical theory to be called Markovian\textemdash or,
perhaps more precisely, \emph{dynamically} Markovian\textemdash is
just to say that its dynamical laws feature a rather nice property:
the dynamical laws, combined with the present-moment state of affairs,
determine the state of affairs at the (possibly infinitesimally) next
moment in time, with no dependence on the past except for whatever
is mediated through the present-moment state of affairs. In somewhat
more detail, given only the present-moment configuration or state
of a system described by a Markovian theory, the dynamical laws fix
either the unique configuration or state at the next moment in time,
if the laws are deterministic, or fix a probability distribution over
configurations or states at the next moment in time, if the laws are
stochastic. A physical theory that fails to have this Markov property
is said to be non-Markovian.

If a non-Markovian physical theory can be re-expressed as a Markovian
theory by formally augmenting its configurations or states with a
suitable collection of unobservable variables, then the resulting
theory is called a hidden Markov model. The unobservable variables
added to make the theory look Markovian are said to be latent or hidden
variables, and they commonly have most of the following seven (and
conceivably more) hallmark characteristics:
\begin{enumerate}
\item (Abstraction) They tend to be conceptually abstract.
\item (Non-uniqueness) They are highly non-unique, in the sense of admitting
a diverse set of mathematical redefinitions.
\item (Unobservability) They are typically unobservable in principle.
\item (Non-spatiality) They have no notion of location in physical space.
\item (Absence of backreaction) They are not backreacted upon by the physical
configurations or states of the system, meaning that they change in
time on their own in a manner that is completely insensitive to the
goings-on of the physical configurations or states of the system.
\item (Multivariateness) They often encompass many degrees of freedom, depending
on how difficult it is to turn the non-Markovian theory into a hidden
Markov model.
\item (Contingency) They feature contingent patterns of time evolution depending
on correspondingly contingent initial conditions.
\end{enumerate}

Hidden Markov models first appeared in the research literature in
the 1960s. They were originally called ``probabilistic functions
of Markov chains'' (Baum, Petrie 1966; Baum, Eagon 1967; Baum, Petrie,
Soules, Weiss 1970; Baum 1972)\nocite{BaumPetrie:1966sifpfofsmc,BaumEagon:1967aiwatsefpfompatamfe,BaumPetrieSoulesWeiss:1970amtoitsaopfomc,Baum:1972aiaamtiseopfoamp}.
They only got their more modern name, ``hidden Markov models,''
in 1970 in a set of unpublished lectures by Lee Neuwirth (Neuwirth
1970; Neuwirth, Cave 1980; Poritz 1988\nocite{Neuwirth:1970ul,CaveNeuwirth:1980hmmfe,Poritz:1988hmmagt}),
and were quickly put to use in statistically modeling speech and language.\footnote{Much of the research on hidden Markov models was classified at the
time, and took place as part of the Institute for Defense Analyses,
where Neuwirth was the deputy director of its Communications Research
Division, despite Neuwirth's strongly anti-war political leanings
(Neuwirth 2009)\nocite{Neuwirth:2009nptvwip11}. The fact that this
research was classified may have further delayed its diffusion into
other research communities, including those working in quantum foundations.
As an interesting historical aside, one of Lee Neuwirth's children
is the actress and singer Bebe Neuwirth, famous for, among other major
roles, playing the character Lillith Sternin on the television sitcoms
\emph{Cheers} and \emph{Frasier}.} More modern examples include $\epsilon$-machines, which have found
widespread use in statistical physics and in the theory of complex
systems (Crutchfield, Young 1989; Travers, Crutchfield 2011)\nocite{CrutchfieldYoung:1989isc,TraversCrutchfield:2011esffss}.

Given that these developments all took place long after quantum theory
and its prominent interpretations had become firmly established in
physics and philosophy, it is hardly surprising that no one thought
to revisit the metaphysical status of quantum states from the standpoint
of hidden Markov models until now.

\subsection{Outline of this Paper\label{subsec:Outline-of-this-Paper}}

To narrow this paper's scope to a manageable degree, the primary focus
will be on pilot-wave theories of fixed numbers of finitely many non-relativistic
particles, like the original de Broglie-Bohm theory, also known as
Bohmian mechanics, which will be reviewed in Section~\ref{sec:Pilot-Wave-Theories}.
On these theories, the quantum state generally takes the form of a
complex-valued wave function $\Psi\left(q,t\right)$ evolving with
the time $t$ according to the Schrödinger equation in the system's
configuration space, each of whose points $q$ denotes a kinematically
allowed \textquoteleft classical\textquoteright{} configuration of
the system. This wave function, also called a pilot wave in the context
of the de Broglie-Bohm theory, then guides or pilots the system's
particles along their trajectories through three-dimensional physical
space according to a precise guiding equation.

Section~\ref{sec:Hidden-Markov-Models} will argue that this pilot-wave
theory can be understood\textemdash and is perhaps best understood\textemdash as
a new kind of hidden Markov model, with the pilot wave  making up
the model's latent or hidden variables. Indeed, as that section of
the paper will show, the pilot wave exhibits all seven smoking-gun
characteristics of latent variables listed above in Subsection~\ref{subsec:Markovian-Theories-and-Hidden-Markov-Models}.

Section~\ref{sec:Further-Challenges-to-the-Ontological-View} will
present additional challenges both to the ontological view of the
wave function, and also to the de Broglie-Bohm pilot-wave theory itself.
These challenges will come from three primary directions: from thinking
about interference effects, from a little-known class of gauge transformations
first introduced by Foldy and Wouthuysen (1950)\nocite{FoldyWouthuysen:1950otdtos12painrl},
and from probing a notion of phase space for the wave function itself
that was defined independently by Strocchi (1966)\nocite{Strocchi:1966ccqm}
and Heslot (1985)\nocite{Heslot:1985qmaact}.

\subsection{Some Terminological Disambiguation\label{subsec:Some-Terminological-Disambiguation}}

Unfortunately, there are a significant number of terminological collisions
between the theory of stochastic processes, the theory of causal modeling,
and quantum theory. Before continuing on to the rest of this paper,
it will be important to disambiguate some of this terminology.

In particular, the terms \textquoteleft model,\textquoteright{} \textquoteleft Markov,\textquoteright{}
and \textquoteleft latent variable\textquoteright{} show up both in
the theory of stochastic processes and in the theory of causal modeling.
Both theories also use similar-looking graphical depictions, called
directed graphs. Meanwhile, the terms \textquoteleft Markov\textquoteright{}
and \textquoteleft hidden variables\textquoteright{} show up both
in the theory of stochastic processes and in quantum theory.

For the theory of stochastic processes, a \textquoteleft model\textquoteright{}
refers to a dynamical process in which one or more random variables
and their probability distributions change with time. In that sense,
a stochastic process generalizes a time series, which consists of
a sequence of numerical values of a particular quantity indexed by
time. By contrast, for the theory of causal modeling, a \textquoteleft model\textquoteright{}
refers to a network of events or random variables connected to each
other with directed causal links.

For the theory of stochastic processes, \textquoteleft Markov\textquoteright{}
should be understood to mean \textquoteleft dynamical Markovianity,\textquoteright{}
referring to the condition that future configurations or states of
the system, or probability distributions over future configurations
or states of the system, are always entirely fixed by just the (possibly
infinitesimally) previous-moment configuration or state of the system
(Milz, Modi 2021)\nocite{MilzModi:2021qspaqnp}. By contrast, for
the theory of causal modeling, \textquoteleft Markov\textquoteright{}
usually refers to the \textquoteleft causal Markov condition,\textquoteright{}
which is the condition that if the probability distribution of a random
variable in a causal model is conditioned on specific values of all
its immediate causal predecessors, or \textquoteleft parents,\textquoteright{}
then that random variable has no correlations with any other random
variables except for any of its causally descendant random variables.
That is, for a causal model satisfying the causal Markov condition,
the parents of a given random variable always screen off all the random
variable's correlations except possibly for correlations with the
random variable's causal descendants. (See, for example, Theorem 1.4.1
in Pearl 2009.)\nocite{Pearl:2009cmrai}

It is possible for a stochastic process to be related to a causal
model for which the stochastic process fails to be dynamically Markovian
while the causal model satisfies the causal Markov condition, and
vice versa. To see why, note that if the configuration of a stochastic
process at some time depends on, say, its predecessors at two previous
times, so that the stochastic process is not dynamically Markovian,
then it may still be possible to construct a causal model in which
those two previous moments in time, treated as parents, screen off
all other correlations, so that the causal model satisfies the causal
Markov condition. Going in the other direction, a dynamically Markovian
process may feature correlations between variables at a single moment
in time that, when written as a causal model, fail to feature the
screening that is required by the causal Markov condition.

To make matters even more confusing, the term \textquoteleft Markov\textquoteright{}
has a subtly different meaning in much of the contemporary research
literature on quantum theory. Conventionally speaking, a quantum system
is said to be Markovian if its present-moment quantum state fully
determines its next quantum state (Milz, Modi 2021)\nocite{MilzModi:2021qspaqnp},
as is the case for closed quantum systems evolving according to unitary
dynamics or for open quantum systems evolving according to the Lindblad
or GKLS equation (Gorin, Kossakowski, Sudarshan 1976; Lindblad 1976)\nocite{GoriniKossakowskiSudarshan:1976cpdsonls,Lindblad:1976gqds}.
Otherwise the quantum system is said to be non-Markovian.

The term \textquoteleft latent\textquoteright{} also merits careful
untangling. For the theory of stochastic processes, a hidden Markov
model is a stochastic process that formally satisfies dynamical Markovianity
due to the augmentation of the system's configurations by additional,
\textquoteleft latent\textquoteright{} variables, where those latent
variables are often entirely unphysical and abstract. By contrast,
for a causal model, \textquoteleft latent structures\textquoteright{}
refer to unobservable components that are needed to ensure the validity
of the causal Markov condition. A causal model that fails to satisfy
the causal Markov condition, due to a missing latent structure, is
usually considered to be an incomplete causal model. Indeed, one major
purpose of causal modeling is to \emph{find} latent structures, which
are usually regarded as physical. (See, for example, Pearl 2009, Subsection
2.9.1, ``On Minimality, Markov, and Stability.'')\nocite{Pearl:2009cmrai}

For a hidden Markov model, latent variables are sometimes called \textquoteleft hidden
variables,\textquoteright{} which should be distinguished from the
notion of hidden variables in quantum theory. In quantum theory, hidden
variables usually refer to any variables other than the quantum state
or wave function that some interpretations regard as representing
necessary ontological components for providing a complete description
of physical reality. For instance, the particles of a pilot-wave theory
are paradigmatic examples of hidden variables according to the quantum-theoretic
meaning of the term. One of the purposes of the present work will
be to argue that the true \textquoteleft hidden variables\textquoteright{}
of quantum theory, at least in the case of pilot-wave interpretations,
are wave functions, in the sense of being best understood as latent
variables of a hidden Markov model.

\section{Pilot-Wave Theories\label{sec:Pilot-Wave-Theories}}

\subsection{Relevant History\label{subsec:Relevant-History}}

Before proceeding to the main arguments of this paper, it will be
helpful to begin with some relevant history, which is covered in more
detail in other work (Barandes 2026b)\nocite{Barandes:2026hdotprotwf}.
The purpose of this history will be to make clear the following four
points about the idea of an ontological wave function or pilot wave
defined in an abstract configuration space rather than in three-dimensional
physical space: 
\begin{itemize}
\item The idea was not a new contribution of David Bohm, Hugh Everett III,
or John Bell, but appeared in Erwin Schrödinger's 1926 papers originally
introducing his wave function.
\item The idea was thoroughly studied and carefully considered in the late
1920s by the major figures responsible for wave-particle duality (Albert
Einstein and Louis de Broglie), the wave function itself (Schrödinger),
and the original pilot-wave theory (de Broglie).
\item To a person, all of them resoundingly, repeatedly, and vociferously
rejected the idea of treating the wave function as ontological, as
did the people responsible for what later became known as the \textquoteleft Copenhagen
interpretation\textquoteright{} (Niels Bohr, Werner Heisenberg, Max
Born, Wolfgang Pauli, and Paul Dirac). Remarkably, denying the physical
reality of the configuration-space wave function was one of the few
things that all the prominent founders of quantum theory agreed on.
\item More favorable attitudes toward the physical or ontological reality
of the wave function showed up starting in the 1950s, and were largely
due to Bohm and Everett, though Schrödinger himself apparently began
warming to the idea around this time as well.
\end{itemize}

In a famous paper in 1905, Albert Einstein proposed that electromagnetic
waves transported their energy in discrete quanta (Einstein 1905)\nocite{Einstein:1905uedeuvdlbhg}.
Partly inspired by this idea, Louis de Broglie, writing in 1923, suggested
that every material particle had an associated \textquoteleft phase
wave\textquoteright{} that propagated in three-dimensional physical
space, maintained phase harmony with any hypothetical periodic processes
internal to the particle, and, in the geometrical-optics limit of
short wavelength, guided or piloted the particle's trajectory along
its rays (de Broglie 1923a\textendash c)\nocite{DeBroglie:1923oeq,DeBroglie:1923qdldei,DeBroglie:1923waq}. 

In de Broglie's 1924 doctoral thesis, which would eventually win him
the 1929 Nobel Prize in physics, he wrote down an early version of
a guiding equation in the form 
\begin{equation}
O_{i}=\frac{1}{h}J_{i},\label{eq:EarlydeBrogleGuidingEqFromThesis}
\end{equation}
 where, in de Broglie's notation, $i$ was a Lorentz index. Here $J_{i}$
was defined to be 
\begin{equation}
J_{i}=m_{0}cu_{i}+e\varphi_{i},\label{eq:deBroglieWorldFourVector}
\end{equation}
 where $m_{0}$ was the particle's proper or rest mass, $c$ was the
speed of light, $e$ was its electric charge, $\varphi_{i}$ were
the electromagnetic gauge potentials (not to be confused with the
notation for the wave's phase function, to be introduced momentarily),
and $u_{i}=dx_{i}/ds$ was the particle's dimensionless four-velocity,
with incremental parameter $ds=\sqrt{c^{2}{dt}^{2}-{dx}^{2}-{dy}^{2}-{dz}^{2}}$.
Meanwhile, $O_{i}$ was given in terms of the phase function $\varphi\left(x,y,z,t\right)$
of the associated wave according to the differential identity 
\begin{equation}
d\varphi=2\pi\sum_{i}O_{i}dx^{i},\label{eq:deBroglieWaveVectorAsSum}
\end{equation}
 so that $O_{i}$ itself, from a more modern standpoint, can be identified
as the wave's four-dimensional wave vector. The spatial parts $\vec O=\left(O_{x},O_{y},O_{z}\right)$
of $O_{i}$ are then the gradient of the phase function $\varphi\left(x,y,z,t\right)$,
up to a reciprocal factor of $2\pi$: 
\begin{equation}
\vec O=\frac{1}{2\pi}\nabla\varphi.\label{eq:deBroglieWaveVectorAsGradient}
\end{equation}
 Hence, in the non-relativistic limit, for which $c\vec u\approx\vec v$,
where $\vec v$ is the particle's ordinary velocity, and assuming
vanishing electromagnetic gauge potentials, $\varphi_{i}=0$, the
primitive guiding equation \eqref{eq:EarlydeBrogleGuidingEqFromThesis}
reduces to 
\begin{equation}
\vec v=\frac{1}{m_{0}}\left(\frac{h}{2\pi}\right)\nabla\varphi.\label{eq:DeBroglieNonRelativisticGuidingEq}
\end{equation}
 Today, the factor in parentheses would be called the reduced Planck
constant $\hbar$ (\textquoteleft h-bar\textquoteright ), as originally
introduced in 1928 by Paul Dirac (Dirac 1928)\nocite{Dirac:1928zqde}:
\begin{equation}
\hbar=\frac{h}{2\pi}.\label{eq:ReducedPlanckConstHbar}
\end{equation}

Partly inspired by de Broglie's phase-wave theory, and partly by Hamilton-Jacobi
theory, Erwin Schrödinger introduced his theory of ``undulatory mechanics''
or ``wave mechanics'' in a series of four foundational papers in
1926 (Schrödinger 1926a\textendash d)\nocite{Schrodinger:1926qae1,Schrodinger:1926qae2,Schrodinger:1926qae3,Schrodinger:1926qae4}.
On this new theory, every quantum system as a whole had a complex-valued
wave function $\psi\left(q,t\right)$ depending on points $q$ in
the system's abstract configuration space and on the time $t$, and
satisfying what is now known as the Schrödinger equation, originally
written as eq. ($4^{\prime\prime})$ in Schrödinger fourth paper (Schrödinger
1926d)\nocite{Schrodinger:1926qae4}: 
\begin{equation}
\Delta\psi-\frac{8\pi^{2}}{h^{2}}V\psi\mp\frac{4\pi i}{h}\frac{\partial\psi}{\partial t}=0.\label{eq:SchrodingerOriginalEq}
\end{equation}
 Here $\Delta$ was a second-order differential operator acting on
the system's configuration space that implicitly involved the masses
of the various particles comprising the system, and $V$ was a potential
function defined on the system's configuration space. The ambiguous
$\mp$ sign appearing in the equation reflected the freedom to work
with either $\psi$ or its complex-conjugate $\bar{\psi}$. Choosing
the positive sign convention, rearranging, and using the definition
of the reduced Planck constant $\hbar=h/2\pi$ from \eqref{eq:ReducedPlanckConstHbar},
the Schrödinger equation takes its more modern-looking form: 
\begin{equation}
i\hbar\frac{\partial\psi}{\partial t}=-\frac{\hbar^{2}}{2}\Delta\psi+V\psi.\label{eq:SchrodingerOriginalEqMoreFamiliarForm}
\end{equation}

As Schrödinger pointed out in the first of these four foundational
papers on undulatory mechanics (Schrödinger 1926a)\nocite{Schrodinger:1926qae1},
his wave function $\psi\left(q,t\right)$ was defined not in three-dimensional
physical space, like de Broglie's phase waves, but in the system's
abstract, generically many-dimensional configuration space. As a consequence,
Schrödinger spent a great deal of effort grappling with the physical
meaning of the wave function in these four papers, in personal correspondence
with colleagues, and in a series of four lectures that he delivered
at the Royal Institution in London in 1928 (Schrödinger 1928)\nocite{Schrodinger:1928flowmdatrilo571a1m1}. 

By the late spring of 1926, Schrödinger had settled on the tentative
idea that the wave function was a physical object of some kind, and
that it manifested itself through its modulus-square $\psi\bar{\psi}=\verts{\psi}^{2}$,
which, in turn, indirectly determined the distribution of electric
charge in three-dimensional physical space. Schrödinger described
this view in the fourth of his foundational 1926 papers, where he
also expressed the speculative idea that the wave function $\psi$
represented the system being in all its kinematical configurations
simultaneously, though in some configurations ``more strongly''
than in others, as a decades-early prefiguring of Hugh Everett's \textquoteleft many
worlds\textquoteright{} interpretation (Schrödinger 1926d)\nocite{Schrodinger:1926qae4}.\footnote{As Everett made clear in his unpublished long-form dissertation in
1956, Schrödinger's later views on the ontology of the wave function
provided inspiration for Everett's own interpretation of quantum theory
(Everett 1956)\nocite{Everett:1956ttotuwf}.}

During the period from 1926 to 1927, Hendrik Lorentz and Albert Einstein
expressed their misgivings with the notion of physical waves propagating
in many-dimensional configuration spaces. Einstein, in particular,
wrote several letters during that period to Lorentz, Max Born, Paul
Ehrenfest, and Arnold Sommerfeld criticizing the idea. Indeed, in
Einstein's famous letter to Born on December 4, 1926, in which Einstein
argued that ``God does not play dice,'' Einstein also included a
complaint about Schrödinger's waves in ``$3n$-dimensional space.''\footnote{This complaint was mistranslated in the canonical English translations
of the Born-Einstein letters (1971)\nocite{EinsteinBornBorn:1971tbelcbaeamahbf1t1wcbmb}.
The crucial ``$n$'' in ``$3n$-dimensional'' was missing (Howard
1990, Barandes 2026b)\nocite{Barandes:2026hdotprotwf}.}

According to Born's statistical hypothesis, presented in a paper by
Born in the summer of 1926, the modulus-square $\verts{\psi\left(q,t\right)}^{2}$
of the wave function represented the probability density for a measurement
of the system's configuration at the time $t$ to yield the value
$q$ (Born 1926)\nocite{Born:1926zqds}. The Born rule planted seeds
of doubt in Schrödinger's mind about the physicality of his wave function.
At the end of his four lectures on wave mechanics in 1928, Schrödinger
said that he had abandoned his earlier view that the wave function
was a physical object, although he moved back toward that view again
later in his life (Schrödinger 1950, 1952a,b)\nocite{Schrodinger:1950wiaep,Schrodinger:1952atqjpi,Schrodinger:1952atqjpii}.

In 1927, de Broglie gave a presentation at the fifth Solvay Conference
laying out a more detailed pilot-wave theory on which the waves associated
with particles guided or piloted them. In a paper that year, he laid
out a version of the theory, now known as his double-solution theory,
that featured two conceptually distinct waves, both satisfying the
same wave equation (de Broglie 1927)\nocite{DeBroglie:1927lmoelsadlmedr}.
One wave was intended to feature a solitonic singularity that represented
a material particle, and the other wave embodied Born's statistical
probability for the particle's location. That 1927 paper was the first
to feature the guiding equation in its more modern-looking form, as
de Broglie's eq. ($26^{\prime}$), 
\begin{equation}
\overrightarrow{v_{M}}=\frac{1}{m_{0}}\overrightarrow{\mathrm{grad}\varphi_{1}},\label{eq:deBroglieGuidingEquationForDoubleSolution}
\end{equation}
 where $\overrightarrow{v_{M}}$ was the particle's velocity, $m_{0}$
was its inertial mass, and $\varphi_{1}$ was the phase function of
the particle's singularity-bearing wave.

Unable to work through the complicated mathematics of his double-solution
theory, de Broglie began working on a second, conceptually distinct
pilot-wave theory in 1928, eventually publishing a detailed description
of this second pilot-wave theory in a 1930 book (de Broglie 1930)\nocite{DeBroglie:1930ialedlmo}.
This alternative pilot-wave theory featured waves and particles as
separate forms of ontology, with the waves again guiding or piloting
the particles along their trajectories in physical space. De Broglie
barely mentioned his double-solution theory in that 1930 book.

However, de Broglie ran into the same conceptual challenges as Schrödinger
in trying to make sense of his theory in the case of multi-particle
systems, for which the associated waves propagated not in three-dimensional
physical space, but in the system's many-dimensional configuration
space, which de Broglie called ``abstract'' and ``fictitious.''
Toward the end of the book, he explained that he had decided to abandon
this second pilot-wave theory as well.

David Bohm independently discovered de Broglie's second pilot-wave
theory in 1951, eventually publishing his telling of the theory in
a pair of 1952 papers submitted for publication simultaneously (Bohm
1952a,b)\nocite{Bohm:1952siqtthvi,Bohm:1952siqtthvii}. In the second
of these two papers, Bohm used his newly formulated theory of decoherence,
as laid out in the final chapters of his 1951 textbook (Bohm 1951,
Chapter 22, ``Quantum Theory of the Measurement Process'')\nocite{Bohm:1951qt},
to clarify how the pilot-wave theory provided a means of resolving
the measurement problem.

In his pair of 1952 papers, Bohm referred to the wave function or
pilot wave as a ``$\psi$-field.'' He also described it in analogy
with the electromagnetic field. However, he said little in these papers
about the $\psi$-field's physical interpretation for multi-particle
systems, although he did briefly acknowledge that the $\psi$-field
would then have to propagate in a many-dimensional space.

In private letters to Wolfgang Pauli in 1951, Bohm explicitly acknowledged
that he had rediscovered de Broglie's second pilot-wave interpretation,
but argued that de Broglie had not taken his theory far enough, and
that Bohm's approach to measurement had resolved several outstanding
limitations of de Broglie's theory. In those letters to Pauli, Bohm
also made an assertive case for regarding the pilot wave as a physical
object, despite its ``polydimensional'' nature (Pauli 1996)\nocite{Pauli:1996scwbehao}.

Einstein and Pauli suggested that Bohm should communicate with de
Broglie. De Broglie immediately wrote a critical paper about the pilot-wave
theory in 1951, emphasizing once again his problem with the idea of
waves propagating in abstract configuration spaces (de Broglie 1951)\nocite{DeBroglie:1951rsltdlop}.
However, Bohm's work rekindled de Broglie's interest in pilot-wave
theories, and de Broglie eventually returned to working on his earlier
double-solution theory from 1927.

Bohm's version of the pilot-wave theory, now known as the de Broglie-Bohm
theory, or Bohmian mechanics, generated new debates over the physicality
of the wave function. Just a few years later, in his unpublished 1956
long-form thesis, Hugh Everett III explicitly cited Bohm's interpretation,
and argued that a system's wave function\textemdash regarded now as
a vector in a Hilbert space, rather than as a function defined in
a configuration space\textemdash was not only ontological, but was
the sole form of ontology in the universe (Everett 1956, 1957a)\nocite{Everett:1956ttotuwf,Everett:1957rsfqm}.
In John Bell's retelling of the de Broglie-Bohm theory, Bell emphasized
the physicality of the pilot wave as well (Bell 1980, 1982)\nocite{Bell:1980dbbdcdseadm,Bell:1982otipw},
even going so far as to write, in italics:
\begin{quotation}
\emph{No one can understand this theory until he is willing to think
of $\psi$ as a real objective field rather than just a \textquoteleft probability
amplitude\textquoteright . Even though it propagates not in $3$-space
but in $3N$-space.} {[}Bell 1981, emphasis in the original{]}\nocite{Bell:1981qmfc}
\end{quotation}
The view that the wave function or pilot wave is physically real or
ontological, despite being defined in a system's configuration space
rather than in physical space, has come to be called \textquoteleft wave-function
realism,\textquoteright{} and continues to generate controversy to
this day (Albert 1996; Lewis 2004; Ney, Albert 2013; Myrvold 2015;
Chen 2019; Wallace 2020; Ney 2021; Ney 2023)\nocite{Albert:1996eqm,AlbertLoewer:1996tossc,Lewis:2004lics,NeyAlbert:2013twfeotmoqm,Myrvold:2015wiaw,Chen:2019ratwf,Wallace:2020awr,Ney:2021twitwfamfqp,Ney:2023tafwfr}.

Pushing back on wave-function realism, Dürr, Goldstein, and Zanghì
have argued that the pilot wave should be understood not as a physical
or ontological object, but as a form of nomology, meaning a feature
of a quantum system's dynamical laws. (See, for example, Dürr, Goldstein,
Zanghì 1996.)\nocite{DurrGoldsteinZanghi:1996bmatmotwf} More will
be said about this nomological view later.

\subsection{General Formulation\label{subsec:General-Formulation}}

It will be useful to lay out the general structure of the de Broglie-Bohm
pilot-wave theory, again also known today as Bohmian mechanics. Let
$\Psi\left(q,t\right)$ denote the complex-valued configuration-space
wave function or pilot wave for a quantum system with $n=3N$ degrees
of freedom and whose configurations are labeled by $q=\left(q_{1},\dots,q_{n}\right)$,
with $t$ denoting the time. Suppose that the configuration-space
wave function satisfies the following schematic version of the Schrödinger
equation \eqref{eq:SchrodingerOriginalEqMoreFamiliarForm}: 

\begin{equation}
i\hbar\,\partial_{t}\Psi=-\frac{\hbar^{2}}{2}\Delta\Psi+V\Psi.\label{eq:SchematicSchrodingerEq}
\end{equation}
 Here $V$ is a potential function defined on the system's configuration
space, $\partial_{t}=\partial/\partial t$ is the partial derivative
with respect to the time $t$, and $\Delta$ is a second-order differential
operator acting on the system's configuration space and having the
general form 

\begin{equation}
\Delta=\sum_{i,j=1}^{n}\partial_{i}\left(\mu_{ij}\partial_{j}\right),\label{eq:SchematicSecondOrderDifferentialOp}
\end{equation}
 where $\mu_{ij}=\mu_{ji}$ is a symmetric array of real-valued functions
of the coordinates (assumed for simplicity to have unit determinant),
$\partial_{i}=\partial/\partial q_{i}$ and $\partial_{j}=\partial/\partial q_{j}$
are the partial derivatives with respect to the respective coordinates
$q_{i}$ and $q_{j}$, and $\Delta$ is understood to act on test
functions $f\left(q,t\right)$ as $\sum_{i,j}\partial_{i}\left(\mu_{ij}\partial_{j}f\right)$.
Writing the wave function in polar form in terms of a real-valued
radial function $R\left(q,t\right)$ and a real-valued phase function
$S\left(q,t\right)/\hbar$, 

\begin{equation}
\Psi\left(q,t\right)=R\left(q,t\right)e^{iS\left(q,t\right)/\hbar},\label{eq:SchematicWaveFunctionPolarDecomposition}
\end{equation}
 the Schrödinger equation \eqref{eq:SchematicSchrodingerEq} breaks
up into a pair of coupled, real-valued equations involving $R$ and
$S$. One then imposes a guiding equation on the velocities $\dot{Q}_{i}\left(t\right)$
based on the values of the gradients $\partial_{j}S$ of the phase
function evaluated on the system's actual trajectory $Q\left(t\right)=\left(Q_{1}\left(t\right),\dots,Q_{n}\left(t\right)\right)$, 

\begin{equation}
\dot{Q}_{i}\left(t\right)={\sum_{j=1}^{n}\mu_{ij}\partial_{j}S}\Big\vert_{Q\left(t\right)},\label{eq:SchematicVelocitiesGuidingEq}
\end{equation}
 where dots denote time derivatives. It follows that the modulus-square
of the wave function, as given by 

\begin{equation}
\rho=\Psi\overline{\Psi}=\verts{\Psi}^{2}=R^{2},\label{eq:SchematicBornRule}
\end{equation}
 satisfies the following continuity equation, 

\begin{equation}
\partial_{t}\rho=-\sum_{i=1}^{n}\partial_{i}J_{i},\label{eq:SchematicProbabilityContinuityEq}
\end{equation}
 where the probability current densities $J_{i}\left(q,t\right)$
are given by  

\begin{equation}
J_{i}=\sum_{j=1}^{n}\hbar\mu_{ij}\im\bar{\Psi}\partial_{j}\Psi=\rho\sum_{j=1}^{n}\mu_{ij}\partial_{j}S,\label{eq:SchematicProbabilityCurrentDensity}
\end{equation}
 as one can show with a short calculation.\footnote{Here is that calculation: 
\begin{align*}
\partial_{t}\rho & =\partial_{t}\left(\Psi\bar{\Psi}\right)=\left(\partial_{t}\Psi\right)\bar{\Psi}+\Psi\left(\partial_{t}\bar{\Psi}\right)\\
 & =\left(-\frac{\hbar}{2i}\Delta\Psi+\cancel{\frac{1}{i\hbar}V\Psi}\right)\bar{\Psi}+\Psi\left(\frac{\hbar}{2i}\Delta\bar{\Psi}-\cancel{\frac{1}{i\hbar}V\bar{\Psi}}\right)=-\sum_{i=1}^{n}\partial_{i}J_{i},
\end{align*}
 where 
\begin{align*}
J_{i} & =\sum_{j=1}^{n}\frac{\hbar}{2i}\mu_{ij}\left(\bar{\Psi}\partial_{j}\Psi-\Psi\partial_{j}\bar{\Psi}\right)=\sum_{j=1}^{n}\hbar\mu_{ij}\im\bar{\Psi}\partial_{j}\Psi\\
 & =\sum_{j=1}^{n}\hbar\mu_{ij}\im Re^{-iS/\hbar}\partial_{j}\left(Re^{iS/\hbar}\right)=\rho\sum_{j=1}^{n}\mu_{ij}\partial_{j}S.
\end{align*}
} One can then write the guiding equation \eqref{eq:SchematicVelocitiesGuidingEq}
in the equivalent form 

\begin{equation}
\dot{Q}_{i}\left(t\right)=\left.\frac{J_{i}}{\rho}\right\vert _{Q\left(t\right)}=\frac{J_{i}\left(Q\left(t\right),t\right)}{\rho\left(Q\left(t\right),t\right)}.\label{eq:SchematicGuidingEquationAsRatioCurrentDensityToProbabilityDensity}
\end{equation}

These equations guarantee a crucial property called \textquoteleft equivariance,\textquoteright{}
which ensures that if the system's initial probability density is
$\rho\left(q,t_{0}\right)=\verts{\Psi\left(q,t_{0}\right)}^{2}$ at
an initial time $t_{0}$, a condition called the \textquoteleft quantum
equilibrium hypothesis,\textquoteright{} then the system's probability
density at all later times $t$ will continue to coincide with $\rho\left(q,t\right)=\verts{\Psi\left(q,t\right)}^{2}$,
in keeping with the Born rule.

Notice that the final version \eqref{eq:SchematicGuidingEquationAsRatioCurrentDensityToProbabilityDensity}
of the guiding equation does not lead to divide-by-zero errors, because,
by construction, the system has zero probability of ever having an
actual configuration $Q\left(t\right)$ at any time $t$ such that
$\rho\left(Q\left(t\right),t\right)=0$. If the system approaches
regions of its configuration space for which the probability density
$\rho\left(q,t\right)$ gets very small, then the system's velocities
$\dot{Q}_{i}\left(t\right)$ become highly unstable, typically driving
the system away from those regions of its configuration space.

\section{Hidden Markov Models\label{sec:Hidden-Markov-Models}}

\subsection{The Shoemaker Universe\label{subsec:The-Shoemaker-Universe}}

It will be useful now to explain what hidden Markov models are, before
establishing their relationship with the de Broglie-Bohm pilot-wave
theory. A good starting place will be a simple but profound thought
experiment.

In a 1969 paper, Sydney Shoemaker presented an argument intended to
establish the conceivability of durations of time without physical
change (Shoemaker 1969)\nocite{Shoemaker:1969twc}. Following Shoemaker,
one imagines a hypothetical universe that is entirely static and empty
except for three civilizations $A$, $B$, and $C$ that live in three
separate solar systems far apart in space, but still close enough
together that sentient beings in each solar system can see the other
civilizations through telescopes. Each civilization uses the same
notions of days and years, with 365 days making up one year. Furthermore,
the Shoemaker universe features some rather unusual behavior:
\begin{itemize}
\item Every three years, civilizations $B$ and $C$ see civilization $A$
suddenly pause completely for one full year, after which civilization
$A$ unpauses and then proceeds normally, with the inhabitants of
civilization $A$ having no awareness of the paused year.
\item Every four years, civilizations $A$ and $C$ similarly see civilization
$B$ pause for one full year before unpausing again.
\item Every five years, civilizations $A$ and $B$ see civilization $C$
pause for one full year before unpausing again.
\end{itemize}
The inhabitants of the Shoemaker universe would seem to be quite reasonable
if they claimed that every $3\times4\times5=60$ years, their entire
universe paused for a full year, even though no physical changes would
have occurred during that paused year, and the inhabitants would not
be able to obtain any direct empirical evidence that such a year-long
pause had happened.

Accepting this view for the sake of argument, what could the dynamical
laws of the Shoemaker universe be like? The laws would seem to be
manifestly non-Markovian, because the state of the Shoemaker universe
at the end of Day 1 of the 60th year looks just like the state of
the Shoemaker universe at the end of Day 365 of the 60th year, and
yet the state at the beginning of Day 2 is still paused, whereas the
state at the beginning of Day 366 looks like a physical change has
taken place. It is as though, according to the dynamical laws of the
Shoemaker universe, the state of the universe at the beginning of
Day 366 has somehow been determined by the state of the universe a
year in the past. Indeed, Shoemaker said that this universe exhibited
``action at a temporal distance.''

If one is uncomfortable with non-Markovian laws of nature like these,
then there is a simple solution that is always available. One can
turn the Shoemaker universe into a hidden Markov model by positing
the existence of a latent variable $Q\left(t\right)$ that grows steadily
with time, perhaps according to the simple differential equation 
\begin{equation}
\frac{dQ\left(t\right)}{dt}=1.\label{eq:ShoemakerLatentVariableEqOfMotion}
\end{equation}
 Then the dynamical laws of the Shoemaker universe can check on the
value of $Q\left(t\right)$ along with the state of the universe at
the present-time $t$ in order to determine the state of the universe
at the infinitesimally next moment in time $t+dt$. If the latent
variable $Q\left(t\right)$ has reached a value corresponding to the
end of Day 365 of a year that is a multiple of 60, then the laws should
prescribe that the next moment should exhibit ordinary physical change
in the three civilizations.

The latent variable $Q\left(t\right)$ features six of the seven telltale
characteristics listed earlier, in Subsection~\ref{subsec:Markovian-Theories-and-Hidden-Markov-Models} 
\begin{enumerate}
\item It is a conceptually abstract variable.
\item It is non-unique and can be deformed by various mathematical transformations
and redefinitions. Indeed, even its differential equation \eqref{eq:ShoemakerLatentVariableEqOfMotion}
could easily be altered.
\item It is unobservable in principle.
\item It has no location in physical space.
\item It is not backreacted upon by the physical material in the Shoemaker
universe.
\item It is very simple, \emph{in violation} of the generically complicated
nature of latent variables.
\item It has contingent time evolution depending on correspondingly contingent
initial conditions, in the sense that its 60-year \textquoteleft internal
clock\textquoteright{} conceivably could have started with various
initial values.
\end{enumerate}

Suppose that one is committed to interpreting $Q\left(t\right)$ as
more than just a formal latent variable. Then one could, in principle,
either take $Q\left(t\right)$ to be an exotic part of the ontology
of the Shoemaker universe, or one could choose to regard $Q\left(t\right)$
as part of the nomology or dynamical laws themselves. However, there
does not seem to be any knock-down argument for favoring the ontological
view over the nomological view, or vice versa. 

More to the point, there does not appear to be a good argument that
confronting this ontological-nomological fork is obligatory. Indeed,
there is a perfectly reasonable alternative way to understood $Q\left(t\right)$:
treat it as a latent variable in a hidden Markov model for a fundamentally
non-Markovian universe. This alternative view only becomes available
when one is aware of the concept of a hidden Markov model, which Shoemaker
almost certainly could not have known about in 1969.

\subsection{Continuous Stochastic Processes\label{subsec:Continuous-Stochastic-Processes}}

As a more complicated but also much more relevant example, one can
consider a continuous stochastic process of a very general form. To
be precise, suppose that a given system with configurations labeled
smoothly by $q=\left(q_{1},\dots,q_{n}\right)$ has a time-dependent
probability density $\rho\left(q,t\right)$ that is a smooth function
of the coordinates $q_{1},\dots,q_{n}$ and the time $t$, where the
total number $n$ of degrees of freedom is assumed to be finite. This
stochastic process will generically be non-Markovian.

Introduce a \textquoteleft field\textquoteright{} $r\left(q,t\right)$
on the system's configuration space according to the differential
equation 
\begin{equation}
\frac{\partial\left(r^{2}\left(q,t\right)\right)}{\partial t}=\frac{\partial\rho\left(q,t\right)}{\partial t},\label{eq:GeneralHMMDiffEqRadialFunction}
\end{equation}
  and define a corresponding set of probability current densities
$J_{i}\left(q,t\right)$ according to\footnote{The time derivative here needs to be a partial derivative because
the integration yields a function of $n+1$ variables.} 
\begin{equation}
J_{i}\left(q_{1},\dots,q_{i},\dots,q_{n},t\right)=-c_{i}\frac{\partial}{\partial t}\int_{a_{i}}^{q_{i}}dq_{i}^{\prime}\,r^{2}\left(q_{1},\dots,q_{i}^{\prime},\dots,q_{n},t\right),\label{eq:GeneralHMMCurrentDensities}
\end{equation}
 where $a_{i}$ are arbitrary constants and where $c_{i}$ are constants
satisfying the summation identity 
\begin{equation}
\sum_{i=1}^{n}c_{i}=1.\label{eq:GeneralHMMCoefficientsSumToOne}
\end{equation}
  It follows that if the following guiding equation is imposed on
the velocities $\dot{Q}_{i}\left(t\right)$ of the system's actual
trajectory $Q\left(t\right)=\left(Q_{1}\left(t\right),\dots,Q_{n}\left(t\right)\right)$,
\begin{equation}
\dot{Q}_{i}\left(t\right)=\left.\frac{J_{i}}{r^{2}}\right\vert _{Q\left(t\right)}=\frac{J_{i}\left(Q\left(t\right),t\right)}{r^{2}\left(Q\left(t\right),t\right)},\label{eq:GeneralHMMGuidingEq}
\end{equation}
 then the system will obey equivariance, in the sense that the identification
\begin{equation}
\rho\left(q,t\right)=r^{2}\left(q,t\right)\label{eq:GeneralHMMRadialFuncSqEqProbabilityDensity}
\end{equation}
 will consistently hold at future times if it holds at any initial
time, in accordance with the continuity equation 
\begin{equation}
\frac{\partial\rho}{\partial t}=-\sum_{i=1}^{n}\frac{\partial J_{i}}{\partial q_{i}},\label{eq:GeneralHHMContinuityEq}
\end{equation}
 which is ensured by the formula \eqref{eq:GeneralHMMCurrentDensities}
for the probability current densities.

The result is then a \emph{deterministic} hidden Markov model for
the continuous stochastic process that is also, furthermore, a pilot-wave
theory. The values of the \textquoteleft field\textquoteright{} $r\left(q,t\right)$
at all points $q$ in the system's configuration space play the role
of both the latent or hidden variables for the hidden Markov model,
and $r\left(q,t\right)$ as a whole is also a pilot wave. 

Notice that $r\left(q,t\right)$, again regarded as an infinite distribution
of variables at all points $q$ in the system's configuration space,
exhibits all seven of the smoking-gun characteristics of latent variables
listed in Subsection~\ref{subsec:Markovian-Theories-and-Hidden-Markov-Models}: 
\begin{enumerate}
\item It is an abstract function defined over the system's abstract configuration
space.
\item It is non-unique in a variety of ways. For example, one can multiply
it by an arbitrary constant without affecting the guiding equation
\eqref{eq:GeneralHMMGuidingEq} or the equivariance property \eqref{eq:GeneralHMMRadialFuncSqEqProbabilityDensity}.
One can also modify the definition of the current densities \eqref{eq:GeneralHMMCurrentDensities}
and therefore the guiding equation by changing the constants $c_{i}$
or by adding terms with vanishing divergence. (See, for example, Deotto,
Ghirardi 1998, which will come up again later.)\nocite{DeottoGhirardi:1998bmr}
\item It is unobservable in principle, because the only given configurations
are those of the original system.
\item It has no notion of location in physical space.
\item It is not backreacted upon by the system's configuration.
\item It encompasses an infinite number of its own degrees of freedom, one
at each point $q$ in the system's configuration space, even if the
original system has only a finite number $n$ of degrees of freedom.
\item It has its own initial conditions, which should satisfy $r^{2}\left(q,t_{0}\right)=\rho\left(q,t_{0}\right)$
at the initial time $t_{0}$.
\end{enumerate}
Notice that the infinite collection of variables that make up $r\left(q,t\right)$
therefore constitute an even better example of latent variables than
the latent variable $Q\left(t\right)$ for the Shoemaker universe
from Subsection~\ref{subsec:The-Shoemaker-Universe} because $Q\left(t\right)$,
by contrast, did not properly exhibit the sixth listed characteristic
(multivariateness).

Looking back at the de Broglie-Bohm pilot-wave theory from Subsection~\ref{subsec:General-Formulation},
one sees a remarkable and telling resemblance. Indeed, it is evident
that the de Broglie-Bohm pilot-wave $\Psi\left(q,t\right)$ exhibits
all seven of these hallmark characteristics of latent variables as
well. The lack of backreaction, or ``back action,'' in particular,
is a widely acknowledged property of the de Broglie-Bohm pilot wave
(Dürr, Goldstein, Zanghì 1996)\nocite{DurrGoldsteinZanghi:1996bmatmotwf}.\footnote{One can, of course, introduce backreactions on the de Broglie-Bohm
pilot wave if one wishes, but the point is that putting in backreactions
by hand is unnecessary.}

One should therefore take seriously the idea that the de Broglie-Bohm
pilot-wave theory is nothing more or less than a deterministic hidden
Markov model of the same general kind as the one constructed here,
with the pilot wave comprising just another class of latent variables.
Indeed, the very facts that the construction laid out in this subsection
was so simple, so general, so non-unique, and so inaccessible to empirical
falsification, suggest that it is essentially just a parlor trick\textemdash an
example of a collapse into triviality\textemdash and not a fundamental
statement about physics.

This new deterministic hidden Markov model even gets the measurement
process to work out correctly, and in a manner very similar to how
the de Broglie-Bohm pilot-wave theory handles measurements (Bohm 1952b)\nocite{Bohm:1952siqtthvii}.
Suppose that the system's overall configuration encompasses a measuring
device containing a \textquoteleft macroscopically large\textquoteright{}
number of its own degrees of freedom, and suppose that the measuring
device has, say, two distinct possible outcomes, each of which corresponds
to a macroscopically distinct rearrangement of its pre-measurement
degrees of freedom. It follows that if the stochastic process is capable
of accommodating measurements in the first place, and if the overall
system goes through a measurement process, then the final probability
density should take the form $\rho\left(q,t\right)=\rho_{1}\left(q,t\right)+\rho_{2}\left(q,t\right)$,
where the two terms correspond to the two macroscopically distinct
measurement outcomes and therefore have non-overlapping support on
the overall system's configuration space. The system's actual configuration
$Q\left(t\right)$ at the time $t$ will belong to just one of these
two regions of support, and so $\rho\left(Q\left(t\right),t\right)$
will reduce to just one of the two terms $\rho_{i}\left(Q\left(t\right),t\right)$,
meaning that only that term $\rho_{i}\left(Q\left(t\right),t\right)$
will show up in the guiding equation \eqref{eq:GeneralHMMGuidingEq}.
By continuity, only that term $\rho_{i}\left(Q\left(t\right),t\right)$
will make a difference to the system's actual trajectory $Q\left(t\right)$
for at least some period of future time. (Beyond that time, there
can, in principle, be \textquoteleft interference effects,\textquoteright{}
but the same is also possible for the de Broglie-Bohm theory in the
case of re-coherence.)

As was the case for the latent variable $Q\left(t\right)$ for the
Shoemaker universe, one is free to claim that the \textquoteleft field\textquoteright{}
$r\left(q,t\right)$ is ontological, or even nomological (meaning
a time-dependent part of the system's dynamical laws), provided that
one is willing to grapple with making sense of ontological objects
in a many-dimensional configuration space, or nomological ingredients
with arbitrarily complicated time dependence and contingent initial
conditions. However, given the arguably more fitting option of understanding
$r\left(q,t\right)$ as a collection of latent variables that are
part of a hidden Markov model, one is no longer obligated\textemdash nor
perhaps even well-advised\textemdash to insist on interpreting $r\left(q,t\right)$
as ontological or nomological.

Crucially, the same option is arguably available for the wave function
or pilot wave $\Psi\left(q,t\right)$ for the de Broglie-Bohm theory.
Why not interpret the wave function as a collection of latent variables
in a hidden Markov model, given its close resemblance to $r\left(q,t\right)$
and the fact that it features all seven smoking-gun characteristics
of a latent variable? If it walks like a duck, and quacks like a duck,
and satisfies \emph{five additional} \emph{characteristics} of a duck,
then what should one conclude? 

\section{Further Challenges to the Ontological View\label{sec:Further-Challenges-to-the-Ontological-View}}

\subsection{Interference Effects\label{subsec:Interference-Effects}}

Beyond the discussion presented so far, there are still other challenges
to the ontological view of the wave function or pilot wave. One place
to start is with questions surrounding the famous interference effects
of quantum theory.

Actually, far from being seen as \emph{undermining} the thesis that
the wave function is ontological, the appearance of interference effects
in various experiments is often held up as evidence \emph{in favor}
of the ontological thesis. It is therefore worth taking a moment to
explain how interference shows up in these sorts of experiments, and
why they are not definitive evidence for an ontological interpretation
of the wave function. A key example is the well-known double-slit
experiment. (See, for example, Feynman, Leighton, Sands 1965, Volume
3, Chapter 1; or Heisenberg 1958, Chapter III. Note that the actual
experiment was not performed with electrons until the work of Jönsson
in 1961.)\nocite{FeynmanLeightonSands:1965tflopv3,Heisenberg:1958paptrims,Jonsson:1961eamkhf,Jonsson:1974edams} 

In the double-slit experiment, particles are sent, \emph{one at a
time}, toward a wall with two small slits or holes in it. The slits
are close together, but are still significantly farther apart than
their individual widths. Far away, on the other side of the wall,
is a detection screen that identifies the precise landing site of
each particle that arrives there. The idea is to collect detailed
statistics on the individual landing sites so that one can construct
an overall histogram.

Crucially, on each successful run of the experiment, there is always
only a single landing site on the detection screen. Obtaining any
statistical patterns among the landing sites therefore requires running
the experiment many times.

If the particles are classical objects, like small stones, then, after
many runs of the experiment, the histogram consists of a blend of
two normal or Gaussian distributions of dots, where each Gaussian
distribution is lined up with one of the two slits.

By contrast, if the experiment is carried out with electrons, then,
again after many runs of the experiment, the histogram appears to
show peaks and valleys of dots, in a manner that looks just like the
pattern of constructive and destructive interference of waves propagating
through the two slits. To be clear, no wavelike interference pattern
is seen on any one run of the experiment\textemdash it is only after
many runs of the experiment that the histogram of landing sites shows
such a pattern of dots.

The textbook explanation for this interference-like pattern of dots
is that each electron is\textemdash or is guided by\textemdash a wave
propagating in three-dimensional physical space, just as de Broglie
originally imagined in his 1923 papers. However, the moment one considers
doing the experiment with multi-electron systems on each run, even
if electrical repulsive effects can be ignored, this intuitive picture
breaks down, because then it becomes salient that the wave function
or pilot wave propagates not in three-dimensional physical space,
but in the system's many-dimensional configuration space. That is,
one encounters precisely the sort of confusing picture that led all
the founders of quantum theory to abandon the ontological view of
the wave function, as outlined in Subsection~\ref{subsec:Relevant-History}
and described more extensively in other work (Barandes 2026b)\nocite{Barandes:2026hdotprotwf}.
Rather than evidence in favor of the ontological view of the wave
function, interference, at least when involving multiple particles
at a time, undermines that view.

Ultimately the wave function is not observable, and is not visualized
in these experiments. What is visualized is the set of landing sites
of the electrons, and whatever one's interpretation of the wave function,
all empirically adequate formulations of quantum theory agree on those
landing sites. On traditional versions of the de Broglie-Bohm pilot-wave
theory, those landing sites are explained by an ontological wave function
interfering with itself and physically guiding the particles preferentially
to regions of the detection screen where the wave function exhibits
constructive interference. On nomological versions of the pilot-wave
theory, the particles are simply obeying complicated dynamical laws,
where those laws have contingent initial conditions and intricate
time dependence of their own.

Alternatively, on the view that the wave function is merely a latent
variable of a hidden Markov model, as argued in this paper, the underlying
physical story is non-Markovian, and the landing sites are the indirect
result of those non-Markovian laws.

An analogy due to Hugh Everett III may be apt here. In a private letter
to Bryce DeWitt, dated May 31, 1957, Everett wrote:
\begin{quotation}
A crucial point in deciding on a theory is that one does \emph{not}
accept or reject the theory on the basis of whether the basic world
picture it presents is compatible with everyday experience. Rather,
one accepts or rejects on the basis of whether or not the \emph{experience
which is predicted by the theory} is in accord with actual experience.

Let me clarify this point. One of the basic criticisms leveled against
the Copernican theory was that the ``mobility of the earth as a real
physical fact is incompatible with the common sense interpretation
of nature.'' In other words, as any fool can plainly see{[},{]} the
earth doesn't \emph{really} move{[},{]} because we don't experience
any motion. However, a theory which involves the motion of the earth
is not difficult to swallow if it is a complete enough theory that
one can also deduce that no motion will be felt by the earth's inhabitants
(as was possible with Newtonian physics). Thus, in order to decide
whether or not a theory contradicts our experience, it is necessary
to see what the theory itself predicts our experience will be. {[}Everett
1957b, emphasis and parenthetical in the original{]}\nocite{Everett:1957ltbd}
\end{quotation}
Everett included similar comments in the shorter, published version
of his doctoral thesis (Everett 1957a)\nocite{Everett:1957rsfqm}.
Those comments appeared in a footnote added during the proofing stage.
In that footnote, he criticized arguments against his own theory,
writing that such arguments 
\begin{quotation}
{[}...{]} are like the criticism of the Copernican theory that the
mobility of the earth as a real physical fact is incompatible with
the common sense interpretation of nature because we feel no such
motion. In both cases the argument fails when it is shown that the
theory itself predicts that our experience will be what it in fact
is. (In the Copernican case the addition of Newtonian physics was
required to be able to show that the earth's inhabitants would be
unaware of any motion of the earth.) {[}Ibid., parenthetical in the
original{]}\footnote{This Copernican analogy has become a repeated anecdote in the literature
on the Everett interpretation of quantum theory, where it is usually
retold in terms of an encounter between Elizabeth Anscombe and Ludwig
Wittgenstein that Anscombe reported in an introduction to the \emph{Tractatus}
(Anscombe 1959)\nocite{Anscombe:1959itwt}. See, for instance, Coleman
(1994), Wallace (2012), and Carroll (2019).\nocite{Coleman:1994qmf,Wallace:2012temqtattei,Carroll:2019sdhqwateos}}
\end{quotation}
Everett's words here echo a famous statement made by Einstein, as
quoted in Heisenberg's memoir \emph{Physics and Beyond} (Heisenberg
1971)\nocite{Heisenberg:1971pabeac}: ``It is the theory which decides
what we can observe.''

The planets in the night sky might provide an even better analogy. 

Over many nights, some of the planets appear to follow trajectories
that take them on retrograde-prograde \textquoteleft loops.\textquoteright{}
From a modern perspective, one accounts for this strange apparent
motion by appealing to Newton's complicated theory of mechanics, which
features planets \emph{qua} orbs, inertial masses, forces, accelerations,
equations of motion that are slightly non-Markovian because they feature
second-order time derivatives, and relative perspectives between the
orbs.

However, the resulting trajectories in the sky end up being nearly
periodic over sufficiently large stretches of time, so Fourier's theorem
guarantees that each such planet's observed trajectory in the sky
can be expressed in terms of a discrete Fourier series. One can organize
the modes in that Fourier series into a collection of unobservable
\textquoteleft epicycles upon epicycles upon epicycles,\textquoteright{}
in much the same sense that Ptolemy originally imagined his epicycles,
where each epicycle is a perfect circle of fixed radius rotating at
a constant rate. These Fourier epicycles are just formal visualizations
of the modes in a discrete Fourier series, of course, so they are
unobservable in principle. Indeed, they satisfy most of the characteristics
of latent variables from Subsection~\ref{subsec:Markovian-Theories-and-Hidden-Markov-Models},
except that they have spatial locations and are unique, again by Fourier's
theorem. 

One can also add, of course, that the Newtonian model is slightly
non-Markovian, because its equations of motion are second-order in
time derivatives. The Fourier-epicycle model is therefore, in a literal
sense, a hidden Markov model for the Newtonian model, although admittedly
not a unique way to embed Newtonian mechanics into a Markovian framework.
Indeed, the Hamiltonian phase-space framework provides a conceptually
distinct such approach (Barandes 2026a)\nocite{Barandes:2026adaoqtaiiftcn}.

In simple cases, the Fourier epicycles give a much more intuitive
and direct explanation of the retrograde-prograde loops that some
of the planets make in the night sky. With Fourier's theorem in hand,
it follows that all of observational planetary astronomy could, in
principle, be reduced to the marvelously parsimonious axiom that planetary
motion in the sky is nearly periodic, so that the only task of scientific
astronomy would be to calculate the Fourier amplitudes to any desired
precision based on empirical observation. Planets on perfect circles
that rotate at constant rates\textemdash what could be conceptually
simpler than that? Why prefer Newton's much more complicated theory
that removes the unobservable Fourier epicycles? Why not just reify
the Fourier epicycles and treat them as ontological? Why treat the
retrograde-prograde loops in the sky as red herrings, rather than
as clear signals that Fourier epicycles should be taken seriously
as part of the ontology of nature?

One good reason to be suspicious of regarding the Fourier epicycles
as ontological is that they have too much in common with latent variables
in a hidden Markov model, which should make them immediately suspect
as aspects of the ontology of nature. Moreover, beyond the simplest
cases, the Fourier epicycles become so ornate and complicated that
they do not give an intuitive or easily visualizable explanation of
the retrograde-prograde loops anymore. Finally, the motion of the
planets is not exactly periodic, and the solar system features many
other objects, like the moons of the various planets, and comets on
non-periodic hyperbolic orbits, that cannot be accommodated into a
discrete system of Fourier epicycles. This discussion, of course,
puts aside additional complications from relativity.

One should note the resemblances of these problems to some of the
outstanding challenges of the de Broglie-Bohm theory, which features
a pilot wave satisfying all the characteristics of a latent variable
in a hidden Markov model, and has run into well-known difficulties
in being generalized beyond the case of systems consisting of fixed
numbers of finitely many non-relativistic particles. One might then
be within one's rights to arrive at the following conclusion: wave
functions are the epicycles of the modern age.

\subsection{Foldy-Wouthuysen Gauge Transformations\label{subsec:Foldy-Wouthuysen-Gauge-Transformations}}

One of the seven characteristics of a latent variable in a hidden
Markov model, as listed in Subsection~\ref{subsec:Markovian-Theories-and-Hidden-Markov-Models},
is non-uniqueness. As this next part of the present work will show,
the degree of non-uniqueness of the quantum state\textemdash whether
treated as a configuration-space wave function or as a state vector
in a Hilbert space\textemdash turns out to be even more substantial
than for the \textquoteleft field\textquoteright{} $r\left(q,t\right)$
of the generic hidden Markov model constructed in Subsection~\ref{subsec:Continuous-Stochastic-Processes}.
That is, a quantum state is even more like a set of latent variables
than $r\left(q,t\right)$, despite the fact that $r\left(q,t\right)$
was intended by construction to comprise a set of latent variables.

The starting place is a remarkable feature of general quantum systems
that is not widely known. Given a quantum system described in terms
of a Hilbert-space formulation with state vector $\ket{\Psi\left(t\right)}$,
observables $A\left(t\right)$, and a Hamiltonian $H\left(t\right)$,
there exists a highly general family of gauge transformations that
leave all empirical quantities invariant. These gauge transformations
were first identified in 1950 by Foldy and Wouthuysen in the specific
context of relativistic spin-half particles (Foldy, Wouthuysen 1950)\nocite{FoldyWouthuysen:1950otdtos12painrl}.
Brown later wrote them in a more general form in a paper concerned
with questions of objectivity for quantum systems (Brown 1999)\nocite{Brown:1999aooiqm}.

To begin, consider an arbitrary quantum system in its Hilbert-space
formulation. Letting $V\left(t\right)$ be an arbitrary time-dependent
unitary operator (not to be confused with the notation for a potential
function), a Foldy-Wouthuysen gauge transformation is defined by 
\begin{equation}
\left.\begin{aligned} & \ket{\Psi\left(t\right)} &  & \mapsto & \ket{\Psi^{\prime}\left(t\right)} & =V\left(t\right)\ket{\Psi\left(t\right)},\\
 & A\left(t\right) &  & \mapsto & A^{\prime}\left(t\right) & =V\left(t\right)A\left(t\right)V^{\adj}\left(t\right),\\
 & H\left(t\right) &  & \mapsto & H^{\prime}\left(t\right) & =V\left(t\right)H\left(t\right)V^{\adj}\left(t\right)-i\hbar\,V\left(t\right)\partial_{t}V^{\adj}\left(t\right).
\end{aligned}
\qquad\right\} \label{eq:DefAbstractFoldyWouthuysenGaugeTransformation}
\end{equation}
 The time-dependence of the unitary operator $V\left(t\right)$ here
is significant. The gauge transformations defined above are not the
ordinary kinds of \emph{constant} unitary transformations that are
familiar from most textbooks on quantum theory.

Foldy-Wouthuysen gauge transformations preserve all expectation values
$\expectval{A\left(t\right)}=\bra{\Psi\left(t\right)}A\left(t\right)\ket{\Psi\left(t\right)}$,
so they leave all the empirical predictions of quantum theory exactly
unchanged. Note that Foldy-Wouthuysen gauge transformations do not
produce a \emph{new} quantum system or theory, but merely yield a
new \textquoteleft gauge\textquoteright{} for the \emph{same} quantum
system or theory.

Notice also that the Hamiltonian $H\left(t\right)$ does not transform
like an observable.\footnote{It may come as a surprise to learn that the Hamiltonian is not an
observable. Is it not the case that for a single non-relativistic
particle of mass $m$, position $q$, momentum $p$, and potential
$V\left(q\right)$, the Hamiltonian $H$ is given by $p^{2}/2m+V\left(q\right)$?
The reply is that although the arithmetic combination of observables
$p^{2}/2m+V\left(q\right)$ is indeed gauge-invariant, the Hamiltonian
$H$ will not remain equal to that specific combination of observables
under Foldy-Wouthuysen gauge transformations.} Instead, the transformation law for the Hamiltonian $H\left(t\right)$
has precisely the same form as the transformation law for a non-Abelian
gauge connection. (For a pedagogical treatment of non-Abelian gauge
theories, see, for example, Peskin, Schroeder 1999; Weinberg 1996)\nocite{PeskinSchroeder:1995iqft,Weinberg:1996tqtfii}.
Indeed, if one regards a time-evolving quantum system as a bundle
of identical Hilbert spaces fibered over a one-dimensional manifold
representing the time $t$, then a Foldy-Wouthuysen gauge transformation
can be understood as an independent unitary rotation by $V\left(t\right)$
of the Hilbert-space fiber at each time $t$, with the Hamiltonian
$H\left(t\right)$ serving as the gauge connection. Indeed, one can
re-express the Hilbert-space Schrödinger equation 
\begin{equation}
i\hbar\frac{\partial\ket{\Psi\left(t\right)}}{\partial t}=H\left(t\right)\ket{\Psi\left(t\right)}\label{eq:SchrodingerEq}
\end{equation}
 in the manifestly gauge-covariant form 
\begin{equation}
\mathcal{D}_{t}\ket{\Psi\left(t\right)}=0,\label{eq:FWGaugeCovSchrodingerEq}
\end{equation}
 where the Foldy-Wouthuysen gauge-covariant derivative $\mathcal{D}_{t}$
is defined to be 
\begin{equation}
\mathcal{D}_{t}\defeq\frac{\partial}{\partial t}+\frac{i}{\hbar}H\left(t\right),\label{eq:DefFWGaugeCovDerivative}
\end{equation}
 in much the same way as for a non-Abelian gauge theory.

The existence of Foldy-Wouthuysen gauge transformations should raise
doubts about the viability of any \textquoteleft ontological-monistic\textquoteright{}
interpretation of quantum theory (in the terminology of Subsection~\ref{subsec:The-Interpretation-of-the-Quantum-State})
that asserts that the ontology of nature is exhausted entirely by
abstract state vectors in Hilbert spaces, which are manifestly not
gauge invariant. These doubts ought to be of particular concern for
some versions of the Everett interpretation, including Everett's own
version of his interpretation, which Everett explicitly took to consist
only of the wave function of the universe as its fundamental ontology
(Everett 1956, 1957a; Barandes 2026b)\nocite{Barandes:2026hdotprotwf}. 

The de Broglie-Bohm pilot-wave theory does not fall into the category
of such interpretations, as the theory explicitly posits additional
ontology beyond the wave function, in the form of particles. However,
the existence of Foldy-Wouthuysen gauge transformations also has potential
implications for the de Broglie-Bohm theory, and these implications
will turn out to put serious pressure not only on the ontological
view of the pilot wave, but on the view that the de Broglie-Bohm theory
is well-defined in the first place.

Working in configuration space, consider a quantum system consisting
of $N$ non-relativistic particles with configurations $q$ labeled
by $n=3N$ degrees of freedom $q_{1},\dots,q_{n}$ (usually assumed
to be position coordinates) and Hamiltonian 
\begin{equation}
H\left(q,t\right)=-\frac{\hbar^{2}}{2}\Delta+V\left(q,t\right).\label{eq:DefHamiltonianForConfigSpace}
\end{equation}
 Here the potential function $V\left(q,t\right)$ should not be confused
with the Foldy-Wouthuysen unitary operator $V\left(t\right)$, and
$\Delta$ is the second-order differential operator on the system's
configuration space originally defined in \eqref{eq:SchematicSecondOrderDifferentialOp}:
\[
\Delta=\sum_{i,j=1}^{n}\partial_{i}\left(\mu_{ij}\partial_{j}\right).
\]

Now consider the subclass of Foldy-Wouthuysen gauge transformations
\eqref{eq:DefAbstractFoldyWouthuysenGaugeTransformation} that commute
with all the coordinate operators of the system. In that case, the
time-dependent unitary operator $V\left(t\right)$, when expressed
in the $q$-representation, reduces to a simple phase factor, 
\begin{equation}
e^{i\lambda\left(q,t\right)/\hbar},\label{eq:FWPhaseFactor}
\end{equation}
 where $\lambda\left(q,t\right)$ is a smooth but otherwise arbitrary
function of the configuration $q=\left(q_{1},\dots,q_{n}\right)$
and the time $t$. The wave function $\Psi\left(q,t\right)$ and the
Hamiltonian $H\left(q,t\right)$ then transform respectively as 
\begin{equation}
\left.\begin{aligned} & \Psi &  & \mapsto & \Psi^{\prime} & =e^{i\lambda/\hbar}\Psi,\\
 & H &  & \mapsto & H^{\prime} & =e^{i\lambda/\hbar}He^{-i\lambda/\hbar}-\partial_{t}\lambda,
\end{aligned}
\qquad\right\} \label{eq:DefBohmianFoldyWouthuysenGaugeTransformation}
\end{equation}
 where the first term in $H^{\prime}$ should be understood as acting
on arbitrary test functions $f\left(q,t\right)$ as $e^{i\lambda/\hbar}H\left(e^{-i\lambda/\hbar}f\right)$.

Under this subclass of Foldy-Wouthuysen gauge transformations, the
Born-rule probability density $\rho\left(q,t\right)=\left\vert \Psi\left(q,t\right)\right\vert ^{2}$
in \eqref{eq:SchematicBornRule} is manifestly gauge invariant: 
\begin{equation}
\rho=\verts{\Psi}^{2}\mapsto\rho^{\prime}=\rho.\label{eq:BohmianProbDensityFWGaugeInv}
\end{equation}
 Meanwhile, the radial function $R\left(q,t\right)$ and the phase
function $S\left(q,t\right)$ appearing in the polar decomposition
\eqref{eq:SchematicWaveFunctionPolarDecomposition} of the wave function
transform respectively as 
\begin{align}
R & \mapsto R^{\prime}=R,\label{eq:BohmianRadialFunctionFWGaugeInv}\\
S & \mapsto S^{\prime}=S+\lambda.\label{eq:BohmianRadialFunctionNotFWGaugeInv}
\end{align}
 Because the phase function $S\left(q,t\right)$ is not gauge invariant,
the probability current densities $J_{i}\left(q,t\right)$ defined
in \eqref{eq:SchematicProbabilityCurrentDensity} are also not gauge
invariant. Instead, they transform nontrivially under Foldy-Wouthuysen
gauge transformations according to 
\begin{equation}
J_{i}=\rho\sum_{j=1}^{n}\mu_{ij}\partial_{j}S\mapsto J_{i}^{\prime}=J_{i}+\rho\sum_{j=1}^{n}\mu_{ij}\partial_{j}\lambda.\label{eq:BohmianCurrentDensityNotFWGaugeInv}
\end{equation}
 Unsurprisingly, with the additional structure that turns textbook
quantum mechanics into the de Broglie-Bohm pilot-wave theory, not
all of the Foldy-Wouthuysen gauge invariance of textbook quantum mechanics
is preserved.

Importantly, however, if one restricts further to the smaller subclass
of Foldy-Wouthuysen gauge transformations for which the inhomogeneous
additive term appearing in the transformation formula \eqref{eq:BohmianCurrentDensityNotFWGaugeInv}
for $J_{i}$ has vanishing generalized divergence, in the precise
sense that 
\begin{equation}
\sum_{i=1}^{n}\partial_{i}\left(\rho\sum_{j=1}^{n}\mu_{ij}\partial_{j}\lambda\right)=0,\label{eq:RestrictionConditionFWGaugeFuntionEqZero}
\end{equation}
 then the continuity equation \eqref{eq:SchematicProbabilityContinuityEq}
still holds, as before: 
\begin{equation}
\partial_{t}\rho=-\sum_{i=1}^{n}\partial_{i}J_{i}.\label{eq:BohmianContinuityEqFWGaugeInv}
\end{equation}
 That is, although some of the Foldy-Wouthuysen gauge invariance is
broken by moving from textbook quantum mechanics to the de Broglie-Bohm
theory, there can still be some lingering Foldy-Wouthuysen gauge invariance
that leaves the de Broglie-Bohm theory's self-consistency intact.

To be clear, these Foldy-Wouthuysen gauge transformations do not change
the de Broglie-Bohm pilot-wave theory into some other theory. They
are a family of gauge transformations inherent to the de Broglie-Bohm
theory itself, akin to the electromagnetic gauge transformations of
the Maxwell theory. These Foldy-Wouthuysen gauge transformations also
do not replace the de Broglie-Bohm particles with some other qualitatively
distinct form of ontology.

Nevertheless, the fact that the current densities \eqref{eq:BohmianCurrentDensityNotFWGaugeInv}
are not gauge invariant means that the velocities $\dot{Q}_{i}\left(t\right)$
of the de Broglie-Bohm particles, as given by the guiding equation
in either the forms \eqref{eq:SchematicVelocitiesGuidingEq} or \eqref{eq:SchematicGuidingEquationAsRatioCurrentDensityToProbabilityDensity},
likewise fail to be gauge invariant, but transform nontrivially according
to 
\begin{equation}
\dot{Q}\left(t\right)\mapsto\dot{Q}^{\prime}\left(t\right)=\dot{Q}\left(t\right)+\sum_{j=1}^{n}\mu_{ij}\partial_{j}\lambda.\label{eq:BohmianVelocitiesGuidingEqNotFWGaugeInv}
\end{equation}
 It follows that the resulting trajectories of the de Broglie-Bohm
particles fail to be gauge invariant as well. Nor is there any preferred
or canonical gauge choice of velocities and trajectories from among
all the gauge choices related by \eqref{eq:BohmianVelocitiesGuidingEqNotFWGaugeInv}.
These distinct gauge choices are all indistinguishable at the level
of the empirical output of the de Broglie-Bohm theory.

The ambiguity \eqref{eq:BohmianVelocitiesGuidingEqNotFWGaugeInv}
in the velocities and trajectories of the de Broglie-Bohm particles
first appeared in a paper by Deotto and Ghirardi (1998)\nocite{DeottoGhirardi:1998bmr},
who showed that the guiding equation\textemdash and thus the trajectories
of the de Broglie-Bohm particles\textemdash were empirically underdetermined
in a very explicit sense: one could add an arbitrary divergence-free
term to the current densities without changing the empirical predictions
of the de Broglie-Bohm theory. A new result of the present work is
thus to connect this Deotto-Ghirardi ambiguity with Foldy-Wouthuysen
gauge transformations. More precisely, the present work shows that
the Deotto-Ghirardi ambiguity does not refer to an \emph{ad hoc}
modification of the de Broglie-Bohm theory, but is an expression of
the fact that the velocities and trajectories of the de Broglie-Bohm
theory, along with the pilot wave itself and the probability current
densities, all fail to be gauge-invariant notions under Foldy-Wouthuysen
gauge transformations.

In other physical theories, structures that fail to be invariant under
a relevant notion of gauge transformations, such as the gauge potentials
of Maxwellian electromagnetism or the metric tensor of general relativity,
tend not to be assigned a physical or ontological meaning. Physical
or ontological meaning is instead usually reserved for gauge-invariant
structures. The de Broglie-Bohm theory has a continuously infinite
collection of suitably restricted Foldy-Wouthuysen gauge transformations
that leave the theory internally consistent (in the sense of preserving
the continuity equation) while not leaving invariant the pilot wave,
the probability current densities, the particle velocities, or the
particle trajectories. This fact is potentially a significant problem
for trying to take the pilot wave and the particle trajectories of
the de Broglie-Bohm theory seriously as part of the ontology of nature.

\subsection{Strocchi-Heslot Phase Spaces\label{subsec:Strocchi-Heslot-Phase-Spaces}}

To get a better handle on the meaning of the Foldy-Wouthuysen gauge
transformations described in Subsection~\ref{subsec:Foldy-Wouthuysen-Gauge-Transformations},
it may be illuminating to examine the mathematical structures underlying
the pilot wave or wave function in more detail. In particular, it
will be worthwhile to spend a moment thinking about phase spaces.
The key point is that by involving both particles and a wave function
$\Psi\left(q,t\right)$, the de Broglie-Bohm theory features two distinct
notions of phase space. 

On the one hand, if the de Broglie-Bohm particles are distinguishable
and are finite in number, then they have a finite-dimensional phase
space in essentially the familiar classical sense. Specifically, the
$i$th particle has canonical coordinates $\left(q_{ix},q_{iy},q_{iz}\right)=\left(x_{i},y_{i},z_{i}\right)$
given by the particle's position, and canonical momenta $\left(p_{ix},p_{iy},p_{iz}\right)=\left(m_{i}\dot{x}_{i},m_{i}\dot{y}_{i},m_{i}\dot{z}_{i}\right)$
given by the components of the particle's physical momentum, where
$m_{i}$ is the mass of the $i$th particle and where dots denote
time derivatives. (These particular formulas for the canonical momenta
can be modified if there are electromagnetic fields present.)

On the other hand, the wave function $\Psi\left(q,t\right)$ itself,
as a complex-valued function defined in the space of configurations
$q$ of the particles and depending on the time $t$, consists of
a pair of real-valued functions $X\left(q,t\right)$ and $Y\left(q,t\right)$
that form their own, infinite-dimensional notion of phase space, and
that satisfy a linear system of first-order, classical-looking Hamilton's
equations of motion that descend from the linear Schrödinger equation.
This additional notion of phase space was first identified by Strocchi
(1966)\nocite{Strocchi:1966ccqm} and then independently by Heslot
(1985)\nocite{Heslot:1985qmaact}, and so will be called the Strocchi-Heslot
phase space in the present work.\footnote{Technically speaking, Strocchi and Heslot identified this additional
notion of phase space only for finite-dimensional quantum systems.
Ashtekar and Schilling later generalized Strocchi-Heslot phase spaces
to quantum systems with infinite-dimensional Hilbert spaces (Ashtekar,
Schilling 1999)\nocite{AshtekarSchilling:1999gfoqm}. For more on
Strocchi-Heslot phase spaces, see Barandes (2026a)\nocite{Barandes:2026adaoqtaiiftcn}.} 

One can then understand general Foldy-Wouthuysen gauge transformations
\eqref{eq:DefAbstractFoldyWouthuysenGaugeTransformation} as linear,
time-dependent canonical transformations on the Strocchi-Heslot phase
space of the wave function, rather than as canonical transformations
of any kind on the phase space of the de Broglie-Bohm particles. 

It is important to keep in mind that these are canonical transformations
on the Strocchi-Heslot phase space of the wave function. In particular,
they should not be confused with canonical transformations on the
phase space of the de Broglie-Bohm particles, which were examined
by Stone (1994)\nocite{Stone:1994dtbtstmp}. In particular, Stone's
paper explored how the de Broglie-Bohm pilot wave could be referred
to different choices of canonical coordinates for the particles\textemdash even
by replacing, say, the positions of the particles with their momenta.
The present work, by contrast, leaves the canonical coordinates for
the particles equal to their positions, so that the pilot wave always
refers to those positions.

A classical phase space is a manifold coordinatized by pairs of variables
called canonical coordinates $q_{i}$ and canonical momenta $p_{i}$,
together with a Poisson bracket $\left\{ \slot,\slot\right\} $ defined
for arbitrary functions $f\left(q,p,t\right)$ and $g\left(q,p,t\right)$
on phase space by 
\begin{equation}
\left\{ f,g\right\} =\sum_{i}\left(\frac{\partial f}{\partial q_{i}}\frac{\partial g}{\partial p_{i}}-\frac{\partial g}{\partial q_{i}}\frac{\partial f}{\partial p_{i}}\right).\label{eq:DefPoissonBracket}
\end{equation}
 Once one has specified a Hamiltonian $H\left(q,p,t\right)$, there
is then an associated flow expressible as a system of differential
equations, called Hamilton's equations of motion, that are first-order
in time derivatives: 
\begin{equation}
\begin{aligned}\dot{q}_{i} & =\frac{\partial H}{\partial p_{i}},\\
\dot{p}_{i} & =-\frac{\partial H}{\partial q_{i}},
\end{aligned}
\label{eq:CanonicalEqs}
\end{equation}
 A change of phase-space variables 
\begin{equation}
\begin{aligned}q_{i}^{\prime} & =q_{i}^{\prime}\left(q,p,t\right),\\
p_{i}^{\prime} & =p_{i}^{\prime}\left(q,p,t\right),
\end{aligned}
\label{eq:CanonicalTransformation}
\end{equation}
 is said to be a canonical transformation if all the various Poisson
brackets, when expressed in terms of the new phase-space variables,
maintain their structure, and if the new phase-space variables satisfy
Hamilton's equations for a potentially modified Hamiltonian $H^{\prime}\left(q^{\prime},p^{\prime},t\right)$. 

Each canonical transformation therefore defines an alternative choice
of phase-space variables, and each such choice is called a canonical
frame. An obvious question, then, is whether some canonical frames
are more physical, or more transparently revealing of the system's
ontology, than other canonical frames. 

For example, if one is given only that a specific system has a Hamiltonian
formulation with one canonical coordinate $q$, one canonical momentum
$p$, and Hamiltonian $H=\left(1/2m\right)p^{2}+\left(1/2\right)kq^{2}$
for positive constants $m,k>0$, then the system could be a simple
harmonic oscillator with position $q$, momentum $p$, mass $m$,
and spring constant $k$, or the system could instead be a simple
harmonic oscillator with position $p$, momentum $-q$, mass $1/k$,
and spring constant $1/m$, as follows from a straightforward calculation
of the corresponding canonical transformation. (The minus sign turns
out to be needed for self-consistency.) 

It is easy to come up with much more exotic examples in which different
choices of canonical frame can lead to much more radical differences
in a system's apparent ontology. A pure Hamiltonian formulation of
a classical system turns out to contain very little definite ontological
structure on its own (Curiel 2014)\nocite{Curiel:2014cmiliinh}.

In practice, how does one resolve this ambiguity over the correct
canonical frame? The standard approach is to identify the system's
observable features, along with their physical meanings. Recalling
the simple harmonic oscillator described earlier, if just observing
the system reveals that $q$ identifies its location in physical space,
and $p$ identifies its momentum in the form $m\dot{q}$, then those
observations single out a canonical frame, at least up to less ontological
questions like one's choice of coordinate system for physical space
or one's choice of measurement units.

As explained above, the wave function has its own notion of phase
space, as first identified by Strocchi and Heslot. However, a wave
function, unlike a particle, or even the local value of an electric
field, is not observable. As a consequence, there are no observables
definable from within a Strocchi-Heslot phase space to single out
an ontological canonical frame for the wave function itself. The subclass
of Foldy-Wouthuysen gauge transformations defined in \eqref{eq:DefBohmianFoldyWouthuysenGaugeTransformation}
and satisfying the subsidiary condition \eqref{eq:RestrictionConditionFWGaugeFuntionEqZero}
represent just one kind of canonical transformation on the Strocchi-Heslot
phase space. There are infinitely many other canonical transformations,
leading to infinitely many other canonical frames, each of which describes
a different-looking ontology for the wave function, without any principled
way to single out any one of them. It is not enough to try to identify
the pilot wave with \textquoteleft all these canonical frames\textquoteright{}
in some broad collective or equivalence-class sense, because that
strategy would not even succeed for the elementary case of the simple
harmonic oscillator described above.

\section{Conclusion\label{sec:Conclusion}}

This paper has argued that the de Broglie-Bohm pilot-wave theory is
best understood as a hidden Markov model, and that the configuration-space
wave function, which serves as the pilot wave for the theory, is optimally
interpreted as consisting of a set of latent variables for that hidden
Markov model. On this view, the latent-variable reading is a better
conceptual fit for the wave function than ontological, nomological,
or epistemological readings. 

This paper has argued, moreover, that the de Broglie-Bohm theory suffers
from a potentially catastrophic ill-definiteness under a class of
gauge transformations that do not preserve the pilot wave or the trajectories
of the de Broglie-Bohm particles. For reasons why the invocation of
\textquoteleft weak measurements\textquoteright{} and \textquoteleft weak
values\textquoteright{} do not provide true empirical evidence of
these trajectories, see Barandes (2026c,d)\nocite{Barandes:2026taratpops,Barandes:2026ttwwv}.
As was also mentioned in the present work, similar problems of gauge-invariance
may present a problem for some versions of the Everett interpretation
as well.

Hidden Markov models are ultimately just formal ways of representing
processes whose dynamics are non-Markovian. One might then naturally
wonder whether quantum theory more broadly should be interpreted as
describing non-Markovian physical systems, beyond the simple cases
of non-relativistic systems of particles or the de Broglie-Bohm theory.
In other words, perhaps when one takes physically fundamental non-Markovian
processes and tries to shoehorn them into a Markovian paradigm, the
result is quantum theory, with all its seemingly exotic features.
For arguments along these lines, see Barandes (2025, 2026a)\nocite{Barandes:2025tsqc,Barandes:2026adaoqtaiiftcn}.

\section*{Acknowledgments}

The author would especially like to thank David Albert, Branden Fitelson,
Barry Loewer, and Tim Maudlin for helpful discussions.

\bibliographystyle{1_home_jacob_Documents_Work_My_Papers_2026-Agai___n_in_Bohmian_Mechanics_custom-abbrvalphaurl}
\bibliography{0_home_jacob_Documents_Work_My_Papers_Bibliography_Global-Bibliography}

\providecommand{\etalchar}[1]{$^{#1}$}
\begin{thebibliography}{{Eve}57b}

\bibitem{AlloriGoldsteinTumulkaZanghi:2008otcsobmatgrwt}
V.~Allori, S.~Goldstein, R.~Tumulka, and N.~Zangh{\`\i}.
\newblock ``{On the Common Structure of Bohmian Mechanics and the
  Ghirardi--Rimini--Weber Theory}''.
\newblock {\em The British Journal for the Philosophy of Science}, 59:353--389,
  2008.
\newblock \href {http://arxiv.org/abs/quant-ph/0603027}
  {\path{arXiv:quant-ph/0603027}}, \href {https://doi.org/10.1093/bjps/axn012}
  {\path{doi:10.1093/bjps/axn012}}.

\bibitem{AlbertLoewer:1996tossc}
D.~Z. Albert and B.~Loewer.
\newblock ``{Tails of Schr{\"o}dinger{\rq}s Cat}''.
\newblock In {\em {Perspectives on Quantum Reality: Non-Relativistic,
  Relativistic, and Field-Theoretic}}, pages 81--92. Springer, 1996.
\newblock \href {https://doi.org/10.1007/978-94-015-8656-6_7}
  {\path{doi:10.1007/978-94-015-8656-6_7}}.

\bibitem{Albert:1996eqm}
D.~Z. Albert.
\newblock ``{Elementary Quantum Metaphysics}''.
\newblock In {\em {Bohmian Mechanics and Quantum Theory: An Appraisal}}, pages
  277--284. Springer, 1996.
\newblock \href {https://doi.org/10.1007/978-94-015-8715-0_19}
  {\path{doi:10.1007/978-94-015-8715-0_19}}.

\bibitem{Anscombe:1959itwt}
G.~E.~M. Anscombe.
\newblock {\em {An Introduction to Wittgenstein's Tractatus}}.
\newblock Hutchinson and Company, 1959.

\bibitem{AshtekarSchilling:1999gfoqm}
A.~Ashtekar and T.~A. Schilling.
\newblock ``{Geometrical Formulation of Quantum Mechanics}'', in A.~Harvey,
  editor, {\em {On Einstein's Path: Essays in Honor of Engelbert Schucking}},
  pages 23--65.
\newblock Springer New York, New York, NY, 1999.
\newblock \href {http://arxiv.org/abs/gr-qc/9706069}
  {\path{arXiv:gr-qc/9706069}}, \href
  {https://doi.org/10.1007/978-1-4612-1422-9_3}
  {\path{doi:10.1007/978-1-4612-1422-9_3}}.

\bibitem{Barandes:2025tsqc}
J.~A. Barandes.
\newblock ``{The Stochastic-Quantum Correspondence}''.
\newblock {\em Philosophy of Physics}, 3(1):8, June 2025.
\newblock \href {http://arxiv.org/abs/2302.10778} {\path{arXiv:2302.10778}},
  \href {https://doi.org/10.31389/pop.186} {\path{doi:10.31389/pop.186}}.

\bibitem{Barandes:2026adaoqtaiiftcn}
J.~A. Barandes.
\newblock ``{A Deflationary Account of Quantum Theory and its Implications for
  the Complex Numbers}'', 2026.
\newblock URL: \url{https://philsci-archive.pitt.edu/26048}, \href
  {http://arxiv.org/abs/2602.01043} {\path{arXiv:2602.01043}}.

\bibitem{Barandes:2026hdotprotwf}
J.~A. Barandes.
\newblock ``{Historical Debates over the Physical Reality of the Wave
  Function}''.
\newblock 2026.
\newblock \href {http://arxiv.org/abs/2602.09397} {\path{arXiv:2602.09397}},
  \href {https://doi.org/10.48550/arXiv.2602.09397}
  {\path{doi:10.48550/arXiv.2602.09397}}.

\bibitem{Barandes:2026taratpops}
J.~A. Barandes.
\newblock ``{The ABL Rule and the Perils of Post-Selection}''.
\newblock 2026.
\newblock URL: \url{https://arxiv.org/abs/2602.07402}, \href
  {http://arxiv.org/abs/2602.07402} {\path{arXiv:2602.07402}}, \href
  {https://doi.org/10.48550/arXiv.2602.07402}
  {\path{doi:10.48550/arXiv.2602.07402}}.

\bibitem{Barandes:2026ttwwv}
J.~A. Barandes.
\newblock ``{The Trouble with Weak Values}''.
\newblock 2026.
\newblock \href {http://arxiv.org/abs/2602.09380} {\path{arXiv:2602.09380}},
  \href {https://doi.org/10.48550/arXiv.2602.09380}
  {\path{doi:10.48550/arXiv.2602.09380}}.

\bibitem{Baum:1972aiaamtiseopfoamp}
L.~E. Baum.
\newblock ``{An Inequality and Associated Maximization Technique in Statistical
  Estimation of Probabilistic Functions of a Markov Process}''.
\newblock {\em Inequalities}, 3:1--8, 1972.

\bibitem{BaumEagon:1967aiwatsefpfompatamfe}
L.~E. Baum and J.~A. Eagon.
\newblock ``{An Inequality with Applications to Statistical Estimation for
  Probabilistic Functions of Markov Processes and to a Model for Ecology}''.
\newblock {\em Bulletin of the American Mathematical Society}, 73(3):360, 1967.
\newblock Zbl 0157.11101.
\newblock \href {https://doi.org/10.1090/S0002-9904-1967-11751-8}
  {\path{doi:10.1090/S0002-9904-1967-11751-8}}.

\bibitem{Bell:1980dbbdcdseadm}
J.~S. Bell.
\newblock ``{de Broglie-Bohm, delayed-choice double-slit experiment, and
  density matrix}''.
\newblock {\em International Journal of Quantum Chemistry: Quantum Chemistry
  Symposium}, 14:155--159, 1980.

\bibitem{Bell:1981qmfc}
J.~S. Bell.
\newblock ``{Quantum Mechanics for Cosmologists}''.
\newblock In C.~Isham, R.~Penrose, and D.~Sciama, editors, {\em {Quantum
  Gravity II}}, page 611, January 1981.

\bibitem{Bell:1982otipw}
J.~S. Bell.
\newblock ``{On the Impossible Pilot Wave}''.
\newblock {\em Foundations of Physics}, 12:989--999, October 1982.
\newblock \href {https://doi.org/10.1007/BF01889272}
  {\path{doi:10.1007/BF01889272}}.

\bibitem{Bell:1987atqj}
J.~S. Bell.
\newblock ``{Are There Quantum Jumps?}''.
\newblock pages 41--52, 1987.
\newblock URL: \url{https://cds.cern.ch/record/183184}, \href
  {https://doi.org/10.1017/CBO9780511564253.005}
  {\path{doi:10.1017/CBO9780511564253.005}}.

\bibitem{Bell:1990am}
J.~S. Bell.
\newblock ``{Against `Measurement'}''.
\newblock {\em Physics World}, 3(8):33, August 1990.
\newblock \href {https://doi.org/10.1088/2058-7058/3/8/26}
  {\path{doi:10.1088/2058-7058/3/8/26}}.

\bibitem{Bohm:1951qt}
D.~J. Bohm.
\newblock {\em {Quantum Theory}}.
\newblock Prentice-Hall, Inc., 1951.

\bibitem{Bohm:1952siqtthvi}
D.~J. Bohm.
\newblock ``{A Suggested Interpretation of the Quantum Theory in Terms of
  `Hidden' Variables. I}''.
\newblock {\em Physical Review}, 85(2):166--179, January 1952.
\newblock \href {https://doi.org/10.1103/PhysRev.85.166}
  {\path{doi:10.1103/PhysRev.85.166}}.

\bibitem{Bohm:1952siqtthvii}
D.~J. Bohm.
\newblock ``{A Suggested Interpretation of the Quantum Theory in Terms of
  `Hidden' Variables. II}''.
\newblock {\em Physical Review}, 85(2):180--193, January 1952.
\newblock \href {https://doi.org/10.1103/PhysRev.85.180}
  {\path{doi:10.1103/PhysRev.85.180}}.

\bibitem{Born:1926zqds}
M.~Born.
\newblock ``{Zur Quantenmechanik der Sto{\ss}vorg{\"a}nge (`On the Quantum
  Mechanics of Collision Processes')}''.
\newblock {\em Zeitschrift f{\"u}r Physik}, 37(12):863--867, 1926.
\newblock \href {https://doi.org/10.1007/BF01397477}
  {\path{doi:10.1007/BF01397477}}.

\bibitem{BaumPetrie:1966sifpfofsmc}
L.~E. Baum and T.~Petrie.
\newblock ``{Statistical Inference for Probabilistic Functions of Finite State
  Markov Chains}''.
\newblock {\em The Annals of Mathematical Statistics}, 37(6):1554--1563,
  December 1966.
\newblock \href {https://doi.org/10.1214/aoms/1177699147}
  {\path{doi:10.1214/aoms/1177699147}}.

\bibitem{BaumPetrieSoulesWeiss:1970amtoitsaopfomc}
L.~E. Baum, T.~Petrie, G.~Soules, and N.~Weiss.
\newblock ``{A Maximization Technique Occurring in the Statistical Analysis of
  Probabilistic Functions of Markov Chains}''.
\newblock {\em The Annals of Mathematical Statistics}, 41(1):164--171, February
  1970.
\newblock JSTOR 2239727; MR 0287613; Zbl 0188.49603.
\newblock \href {https://doi.org/10.1214/aoms/1177697196}
  {\path{doi:10.1214/aoms/1177697196}}.

\bibitem{Brown:1999aooiqm}
H.~Brown.
\newblock ``{Aspects of Objectivity in Quantum Mechanics}''.
\newblock In J.~Butterfield and C.~Pagonis, editors, {\em {From Physics to
  Philosophy}}, pages 45--70. Cambridge University Press, 1999.
\newblock URL: \url{https://philpapers.org/rec/BROAOO-2}.

\bibitem{Carroll:2019sdhqwateos}
S.~M. Carroll.
\newblock {\em {Something Deeply Hidden: Quantum Worlds and the Emergence of
  Spacetime}}.
\newblock Dutton, 2019.

\bibitem{Chen:2019ratwf}
E.~K. Chen.
\newblock ``{Realism About the Wave Function}''.
\newblock {\em Philosophy compass}, 14(7):e12611, 2019.
\newblock \href {http://arxiv.org/abs/1810.07010v2}
  {\path{arXiv:1810.07010v2}}, \href {https://doi.org/10.1111/phc3.12611}
  {\path{doi:10.1111/phc3.12611}}.

\bibitem{CaveNeuwirth:1980hmmfe}
R.~L. Cave and L.~P. Neuwirth.
\newblock ``{Hidden Markov Models for English}''.
\newblock In J.~D. Ferguson, editor, {\em {Hidden Markov Models for Speech}},
  Princeton, NJ, Oct. 1980. IDA-CRD.
\newblock URL:
  \url{https://www.cs.sjsu.edu/~stamp/RUA/CaveNeuwirth/index.html}.

\bibitem{Coleman:1994qmf}
S.~Coleman.
\newblock ``{Quantum Mechanics in Your Face}'', 1994.
\newblock Lecture video.
\newblock \href {http://arxiv.org/abs/2011.12671} {\path{arXiv:2011.12671}},
  \href {https://doi.org/10.48550/arXiv.2011.12671}
  {\path{doi:10.48550/arXiv.2011.12671}}.

\bibitem{Curiel:2014cmiliinh}
E.~Curiel.
\newblock ``{Classical Mechanics Is Lagrangian; It Is Not Hamiltonian}''.
\newblock {\em British Journal for the Philosophy of Science}, 65(2):269--321,
  June 2014.
\newblock \href {https://doi.org/10.1093/bjps/axs034}
  {\path{doi:10.1093/bjps/axs034}}.

\bibitem{CrutchfieldYoung:1989isc}
J.~P. Crutchfield, K.~Young, et~al.
\newblock ``{Inferring Statistical Complexity}''.
\newblock {\em Physical review letters}, 63(2):105--108, July 1989.
\newblock \href {https://doi.org/10.1103/PhysRevLett.63.105}
  {\path{doi:10.1103/PhysRevLett.63.105}}.

\bibitem{DeBroglie:1923waq}
L.~de~Broglie.
\newblock ``{Waves and Quanta}''.
\newblock {\em Nature}, 112(2815):540, October 1923.
\newblock \href {https://doi.org/10.1038/112540a0}
  {\path{doi:10.1038/112540a0}}.

\bibitem{DeBroglie:1923oeq}
L.~V. P.~R. de~Broglie.
\newblock ``{Ondes et Quanta}''.
\newblock {\em Comptes Rendus de l'Acad{\'e}mie des Sciences}, 177:507--510,
  September 1923.

\bibitem{DeBroglie:1923qdldei}
L.~V. P.~R. de~Broglie.
\newblock ``{Quanta de Lumi{\`e}re, Diffraction et Interf{\`e}rences}''.
\newblock {\em Comptes Rendus de l'Acad{\'e}mie des Sciences}, 177:548--550,
  September 1923.

\bibitem{DeBroglie:1930ialedlmo}
L.~de~Broglie.
\newblock {\em {Introduction {\`a} l'{\'E}tude de la M{\'e}canique
  Ondulatoire}}.
\newblock Hermann, 1930.
\newblock English translation.

\bibitem{DeBroglie:1951rsltdlop}
L.~V. P.~R. de~Broglie.
\newblock ``{Remarques sur la Th{\'e}orie de l'Onde Pilote}''.
\newblock {\em Comptes Rendus de l'Acad{\'e}mie des Sciences}, 233:614--644,
  September 1951.

\bibitem{DeBroglie:1927lmoelsadlmedr}
L.~{De Broglie}.
\newblock ``{La M{\'e}canique Ondulatoire et la Structure Atomique de la
  Mati{\`e}re et du Rayonnement}''.
\newblock {\em Journal de Physique et le Radium}, 8(5):225--241, 1927.
\newblock \href {https://doi.org/10.1051/jphysrad:0192700805022500}
  {\path{doi:10.1051/jphysrad:0192700805022500}}.

\bibitem{DeottoGhirardi:1998bmr}
E.~Deotto and G.~Ghirardi.
\newblock ``{Bohmian Mechanics Revisited}''.
\newblock {\em Foundations of Physics}, 28(1):1--30, 1998.
\newblock \href {http://arxiv.org/abs/quant-ph/9704021}
  {\path{arXiv:quant-ph/9704021}}, \href
  {https://doi.org/10.1023/A:1018752202576}
  {\path{doi:10.1023/A:1018752202576}}.

\bibitem{DurrGoldsteinZanghi:1996bmatmotwf}
D.~D{\"u}rr, S.~Goldstein, and N.~Zangh{\`\i}.
\newblock ``{Bohmian Mechanics and the Meaning of the Wave Function}''.
\newblock In R.~S. Cohen, M.~A. Horne, and J.~J. Stachel, editors, {\em
  {Experimental Metaphysics: Quantum Mechanical Studies for Abner Shimony,
  Volume 1}}, pages 25--38. Kluwer Academic Publishers, 1996.
\newblock URL: \url{https://arxiv.org/abs/quant-ph/9512031}, \href
  {http://arxiv.org/abs/9512031} {\path{arXiv:9512031}}.

\bibitem{Dirac:1928zqde}
P.~A.~M. Dirac.
\newblock ``{Zur Quantentheorie des Elektrons}''.
\newblock {\em Leipziger Voltr{\"a}ge 1928: Quantentheorie und Chemie}, pages
  85--94, 1928.

\bibitem{Dirac:1930pofm}
P.~A.~M. Dirac.
\newblock {\em {The Principles of Quantum Mechanics}}.
\newblock Oxford University Press, 1st edition, 1930.

\bibitem{EinsteinBornBorn:1971tbelcbaeamahbf1t1wcbmb}
A.~Einstein, M.~Born, and H.~Born.
\newblock {\em {The Born--Einstein Letters: Correspondence between Max \&
  Hedwig Born and Albert Einstein from 1916 to 1955, with commentaries by Max
  Born, translated by Irene Born}}.
\newblock Macmillan, New York, 1971.

\bibitem{EverettDeWittGraham:1973mwiqm}
H.~{Everett III}, B.~S. DeWitt, and N.~Graham.
\newblock {\em {The Many-Worlds Interpretation of Quantum Mechanics}}.
\newblock Princeton University Press, 1973.

\bibitem{Einstein:1905uedeuvdlbhg}
A.~Einstein.
\newblock ``{{\"U}ber einem die Erzeugung und Verwandlung des Lichtes
  betreffenden heuristischen Gesichtspunkt}''.
\newblock {\em Annalen der physik}, 4:132--148, January 1905.
\newblock \href {https://doi.org/10.1002/andp.19053220607}
  {\path{doi:10.1002/andp.19053220607}}.

\bibitem{Everett:1956ttotuwf}
H.~{Everett III}.
\newblock ``{The Theory of the Universal Wave Function}''.
\newblock Unpublished draft Ph.D. thesis (137 pp.), Princeton University, 1956.

\bibitem{Everett:1957ltbd}
H.~Everett.
\newblock ``{Letter to Bryce DeWitt}''.
\newblock Unpublished letter, May 1957.
\newblock Recipient: Bryce DeWitt, Department of Physics, University of North
  Carolina, Chapel Hill.

\bibitem{Everett:1957rsfqm}
H.~{Everett III}.
\newblock ``{{ }`Relative State' Formulation of Quantum Mechanics}''.
\newblock {\em Reviews of Modern Physics}, 29(3):454--462, July 1957.
\newblock \href {https://doi.org/10.1103/RevModPhys.29.454}
  {\path{doi:10.1103/RevModPhys.29.454}}.

\bibitem{FeynmanLeightonSands:1965tflopv3}
R.~P. Feynman, R.~B. Leighton, and M.~Sands.
\newblock {\em {The Feynman Lectures on Physics, Volume 3}}.
\newblock Addison-Wesley, 1965.
\newblock URL: \url{https://www.feynmanlectures.caltech.edu/III_toc.html}.

\bibitem{Fuchs:2010qbpqb}
C.~A. Fuchs.
\newblock ``{QBism, the Perimeter of Quantum Bayesianism}''.
\newblock 2010.
\newblock \href {http://arxiv.org/abs/1003.5209} {\path{arXiv:1003.5209}}.

\bibitem{FoldyWouthuysen:1950otdtos12painrl}
L.~L. Foldy and S.~A. Wouthuysen.
\newblock ``{On the Dirac Theory of Spin 1/2 Particles and Its Non-Relativistic
  Limit}''.
\newblock {\em Physical Review}, 78(1):29--36, April 1950.
\newblock \href {https://doi.org/10.1103/PhysRev.78.29}
  {\path{doi:10.1103/PhysRev.78.29}}.

\bibitem{Gisin:1984qmasp}
N.~Gisin.
\newblock ``{Quantum Measurements and Stochastic Processes}''.
\newblock {\em Physical Review Letters}, 52(19):1657, May 1984.
\newblock \href {https://doi.org/10.1103/PhysRevLett.52.1657}
  {\path{doi:10.1103/PhysRevLett.52.1657}}.

\bibitem{GoriniKossakowskiSudarshan:1976cpdsonls}
V.~Gorini, A.~Kossakowski, and E.~C. G.~S. Sudarshan.
\newblock ``{Completely Positive Dynamical Semigroups of N-Level Systems}''.
\newblock {\em Journal of Mathematical Physics}, 17(5):821--825, May 1976.
\newblock \href {https://doi.org/10.1063/1.522979}
  {\path{doi:10.1063/1.522979}}.

\bibitem{GhirardiRiminiWeber:1986udmms}
G.~Ghirardi, A.~Rimini, and T.~Weber.
\newblock ``{Unified Dynamics for Microscopic and Macroscopic Systems}''.
\newblock {\em Physical Review D}, 34(2):470--491, 1986.
\newblock \href {https://doi.org/10.1103/PhysRevD.34.470}
  {\path{doi:10.1103/PhysRevD.34.470}}.

\bibitem{GriffithsSchroeter:2018iqm}
D.~J. Griffiths and D.~F. Schroeter.
\newblock {\em {Introduction to Quantum Mechanics}}.
\newblock Cambridge University Press, 3rd edition, 2018.

\bibitem{GoldsteinTumulkaZanghi:2012tqfatgf}
S.~Goldstein, R.~Tumulka, and N.~Zangh{\`\i}.
\newblock ``{The Quantum Formalism and the GRW Formalism}''.
\newblock {\em Journal of Statistical Physics}, 149(1):142--201, September
  2012.
\newblock \href {http://arxiv.org/abs/0710.0885v5} {\path{arXiv:0710.0885v5}},
  \href {https://doi.org/10.1007/s10955-012-0587-6}
  {\path{doi:10.1007/s10955-012-0587-6}}.

\bibitem{Heisenberg:1955tdotiotqt}
W.~Heisenberg.
\newblock ``{The Development of the Interpretation of the Quantum Theory}''.
\newblock In W.~Pauli, editor, {\em {Niels Bohr and the Development of
  Physics}}, pages 12--29. Pergamon Press, London, 1955.

\bibitem{Heisenberg:1958paptrims}
W.~Heisenberg.
\newblock {\em {Physics and Philosophy: The Revolution in Modern Science}}.
\newblock Harper \& Brothers Publishers, 1958.

\bibitem{Heisenberg:1971pabeac}
W.~Heisenberg.
\newblock {\em {Physics and Beyond: Encounters and Conversations}}.
\newblock Harper \& Row, 1971.

\bibitem{Heslot:1985qmaact}
A.~Heslot.
\newblock ``{Quantum Mechanics as a Classical Theory}''.
\newblock {\em Physical Review D}, 31(6):1341--1348, March 1985.
\newblock \href {https://doi.org/10.1103/PhysRevD.31.1341}
  {\path{doi:10.1103/PhysRevD.31.1341}}.

\bibitem{Howard:2004witciasim}
D.~Howard.
\newblock ``{Who Invented the ``Copenhagen Interpretation''? A study in
  Mythology}''.
\newblock {\em Philosophy of Science}, 71(5):669--682, 2004.
\newblock \href {https://doi.org/10.1086/425941} {\path{doi:10.1086/425941}}.

\bibitem{HarriganSpekkens:2010eievqs}
N.~Harrigan and R.~W. Spekkens.
\newblock ``{Einstein, Incompleteness, and the Epistemic View of Quantum
  States}''.
\newblock {\em Foundations of Physics}, 40(2):125--157, 2010.
\newblock \href {http://arxiv.org/abs/0706.2661} {\path{arXiv:0706.2661}},
  \href {https://doi.org/10.1007/s10701-009-9347-0}
  {\path{doi:10.1007/s10701-009-9347-0}}.

\bibitem{Jonsson:1961eamkhf}
C.~J{\"o}nsson.
\newblock ``{Elektroneninterferenzen an mehreren k{\"u}nstlich hergestellten
  Feinspalten}''.
\newblock {\em Zeitschrift f{\"u}r Physik}, 161(4):454--474, August 1961.
\newblock \href {https://doi.org/10.1007/BF01342460}
  {\path{doi:10.1007/BF01342460}}.

\bibitem{Jonsson:1974edams}
C.~J{\"o}nsson.
\newblock ``{Electron Diffraction at Multiple Slits}''.
\newblock {\em American Journal of Physics}, 42(1):4--11, January 1974.
\newblock \href {https://doi.org/10.1119/1.1987592}
  {\path{doi:10.1119/1.1987592}}.

\bibitem{Lewis:2004lics}
P.~J. Lewis.
\newblock ``{Life in Configuration Space}''.
\newblock {\em The British journal for the philosophy of science},
  55(4):713--729, 2004.
\newblock \href {https://doi.org/10.1093/bjps/55.4.713}
  {\path{doi:10.1093/bjps/55.4.713}}.

\bibitem{Lindblad:1976gqds}
G.~Lindblad.
\newblock ``{On the Generators of Quantum Dynamical Semigroups}''.
\newblock {\em Communications in Mathematical Physics}, 48(2):119--130, 1976.
\newblock \href {https://doi.org/10.1007/BF01608499}
  {\path{doi:10.1007/BF01608499}}.

\bibitem{Maudlin:1995tmp}
T.~W.~E. Maudlin.
\newblock ``{Three Measurement Problems}''.
\newblock {\em Topoi}, 14(1):7--15, 1995.

\bibitem{MilzModi:2021qspaqnp}
S.~Milz and K.~Modi.
\newblock ``{Quantum Stochastic Processes and Quantum Non-Markovian
  Phenomena}''.
\newblock {\em PRX Quantum}, 2:030201, May 2021.
\newblock \href {http://arxiv.org/abs/2012.01894v2}
  {\path{arXiv:2012.01894v2}}, \href
  {https://doi.org/10.1103/PRXQuantum.2.030201}
  {\path{doi:10.1103/PRXQuantum.2.030201}}.

\bibitem{Myrvold:2015wiaw}
W.~C. Myrvold.
\newblock ``{What is a Wavefunction?}''.
\newblock {\em Synthese}, 192(10):3247--3274, January 2015.
\newblock \href {https://doi.org/10.1007/s11229-014-0635-7}
  {\path{doi:10.1007/s11229-014-0635-7}}.

\bibitem{NeyAlbert:2013twfeotmoqm}
A.~Ney and D.~Z. Albert.
\newblock {\em {The Wave Function: Essays on the Metaphysics of Quantum
  Mechanics}}.
\newblock Oxford University Press, 2013.

\bibitem{Neuwirth:1970ul}
L.~P. Neuwirth.
\newblock ``{Unpublished Lectures}''.
\newblock Unpublished lecture series, 1970.

\bibitem{Neuwirth:2009nptvwip11}
L.~P. Neuwirth.
\newblock {\em {Nothing Personal: The Vietnam War in Princeton 1965--1975}}.
\newblock BookSurge Publishing, Charleston, SC, July 2009.

\bibitem{Ney:2021twitwfamfqp}
A.~Ney.
\newblock {\em {The World in the Wave Function: A Metaphysics for Quantum
  Physics}}.
\newblock Oxford University Press, 2021.
\newblock \href {https://doi.org/10.1093/oso/9780190097714.001.0001}
  {\path{doi:10.1093/oso/9780190097714.001.0001}}.

\bibitem{Ney:2023tafwfr}
A.~Ney.
\newblock ``{Three Arguments for Wave Function Realism}''.
\newblock {\em European Journal for Philosophy of Science}, 13(4):50, October
  2023.
\newblock \href {https://doi.org/10.1007/s13194-023-00554-5}
  {\path{doi:10.1007/s13194-023-00554-5}}.

\bibitem{Pauli:1996scwbehao}
W.~Pauli.
\newblock {\em {Scientific Correspondence with Bohr, Einstein, Heisenberg,
  a.o.}}, Volume~IV.
\newblock Springer, 1996.

\bibitem{PuseyBarrettRudolph:2012rqs}
M.~F. Pusey, J.~Barrett, and T.~Rudolph.
\newblock ``{On the Reality of the Quantum State}''.
\newblock {\em Nature Physics}, 8(6):475--478, 2012.
\newblock \href {http://arxiv.org/abs/1111.3328} {\path{arXiv:1111.3328}},
  \href {https://doi.org/10.1038/nphys2309} {\path{doi:10.1038/nphys2309}}.

\bibitem{Pearl:2009cmrai}
J.~Pearl.
\newblock {\em {Causality: Models, Reasoning and Inference}}.
\newblock Cambridge University Press, 2009.

\bibitem{Poritz:1988hmmagt}
A.~B. Poritz.
\newblock ``{Hidden Markov Models: A Guided Tour}''.
\newblock In {\em {ICASSP}}, Volume~88, pages 7--13, 1988.

\bibitem{PeskinSchroeder:1995iqft}
M.~E. Peskin and D.~V. Schroeder.
\newblock {\em {An Introduction to Quantum Field Theory}}.
\newblock Westview Press, 1995.

\bibitem{Schrodinger:1926qae1}
E.~Schr{\"o}dinger.
\newblock ``{Quantisierung als Eigenwertproblem}''.
\newblock {\em Annalen der Physik}, 79(4):361--376, 1926.
\newblock \href {https://doi.org/10.1002/andp.19263840404}
  {\path{doi:10.1002/andp.19263840404}}.

\bibitem{Schrodinger:1926qae2}
E.~Schr{\"o}dinger.
\newblock ``{Quantisierung als Eigenwertproblem}''.
\newblock {\em Annalen der Physik}, 79(6):489--527, 1926.
\newblock \href {https://doi.org/10.1002/andp.19263840602}
  {\path{doi:10.1002/andp.19263840602}}.

\bibitem{Schrodinger:1926qae3}
E.~Schr{\"o}dinger.
\newblock ``{Quantisierung als Eigenwertproblem}''.
\newblock {\em Annalen der Physik}, 80(13):437--490, 1926.
\newblock \href {https://doi.org/10.1002/andp.19263851302}
  {\path{doi:10.1002/andp.19263851302}}.

\bibitem{Schrodinger:1926qae4}
E.~Schr{\"o}dinger.
\newblock ``{Quantisierung als Eigenwertproblem}''.
\newblock {\em Annalen der Physik}, 81(18):109--139, 1926.
\newblock \href {https://doi.org/10.1002/andp.19263861802}
  {\path{doi:10.1002/andp.19263861802}}.

\bibitem{Schrodinger:1928flowmdatrilo571a1m1}
E.~Schr{\"o}dinger.
\newblock ``{Four Lecture on Wave Mechanics: Delivered at the Royal
  Institution, London, on 5th, 7th, 12th, and 14th March, 1928}''.
\newblock 1928.

\bibitem{Schrodinger:1950wiaep}
E.~Schr{\"o}dinger.
\newblock ``{What is an Elementary Particle?}''.
\newblock {\em Endeavour}, 9(35), July 1950.

\bibitem{Schrodinger:1952atqjpi}
E.~Schr{\"o}dinger.
\newblock ``{Are There Quantum Jumps? Part I}''.
\newblock {\em The British Journal for the Philosophy of science},
  3(10):109--123, August 1952.
\newblock URL: \url{https://doi.org/10.1093/bjps/III.10.109;
  https://www.jstor.org/stable/685552}, \href
  {https://doi.org/10.1093/bjps/III.10.109}
  {\path{doi:10.1093/bjps/III.10.109}}.

\bibitem{Schrodinger:1952atqjpii}
E.~Schr{\"o}dinger.
\newblock ``{Are There Quantum Jumps? Part II}''.
\newblock {\em The British Journal for the Philosophy of science},
  3(11):233--242, November 1952.
\newblock URL: \url{https://doi.org/10.1093/bjps/III.11.233;
  https://www.jstor.org/stable/685266}, \href
  {https://doi.org/10.1093/bjps/III.11.233}
  {\path{doi:10.1093/bjps/III.11.233}}.

\bibitem{Segal:1947otionritroooahs}
I.~E. Segal.
\newblock ``{Irreducible Representations of Operator Algebras}''.
\newblock {\em Bulletin of the American Mathematical Society}, 53:73--88, 1947.
\newblock \href {https://doi.org/10.1090/S0002-9904-1947-08742-5}
  {\path{doi:10.1090/S0002-9904-1947-08742-5}}.

\bibitem{Segal:1947pfgqm}
I.~E. Segal.
\newblock ``{Postulates for General Quantum Mechanics}''.
\newblock {\em Annals of Mathematics}, 48(4):930--948, October 1947.
\newblock \href {https://doi.org/10.2307/1969387} {\path{doi:10.2307/1969387}}.

\bibitem{Shankar:1994pqm}
R.~Shankar.
\newblock {\em {Principles of Quantum Mechanics}}.
\newblock Plenum Press, 2nd edition, 1994.

\bibitem{Shoemaker:1969twc}
S.~Shoemaker.
\newblock ``{Time Without Change}''.
\newblock {\em The Journal of Philosophy}, 66(12):363--381, June 1969.
\newblock URL: \url{https://www.jstor.org/stable/2023892}.

\bibitem{SakuraiNapolitano:2010mqm}
J.~J. Sakurai and J.~J. Napolitano.
\newblock {\em {Modern Quantum Mechanics}}.
\newblock Addison-Wesley, 2nd edition, 2010.

\bibitem{Stone:1994dtbtstmp}
A.~D. Stone.
\newblock ``{Does the Bohm Theory Solve the Measurement Problem?}''.
\newblock {\em Philosophy of Science}, 61(2):250--266, June 1994.
\newblock URL: \url{https://doi.org/10.1086/289798;
  https://www.jstor.org/stable/188211}, \href {https://doi.org/10.1086/289798}
  {\path{doi:10.1086/289798}}.

\bibitem{Strocchi:1966ccqm}
F.~Strocchi.
\newblock ``{Complex Coordinates and Quantum Mechanics}''.
\newblock {\em Reviews of Modern Physics}, 38(1):36--40, 1966.
\newblock \href {https://doi.org/10.1103/RevModPhys.38.36}
  {\path{doi:10.1103/RevModPhys.38.36}}.

\bibitem{TraversCrutchfield:2011esffss}
N.~F. Travers and J.~P. Crutchfield.
\newblock ``{Exact synchronization for finite-state sources}''.
\newblock {\em Journal of Statistical Physics}, 145(5):1181--1201, September
  2011.
\newblock \href {http://arxiv.org/abs/1008.4182} {\path{arXiv:1008.4182}},
  \href {https://doi.org/10.1007/s10955-011-0342-4}
  {\path{doi:10.1007/s10955-011-0342-4}}.

\bibitem{vonNeumann:1932mgdq}
J.~von Neumann.
\newblock {\em {Mathematische Grundlagen der Quantenmechanik}}.
\newblock Berlin: Springer, 1932.

\bibitem{Wallace:2012temqtattei}
D.~Wallace.
\newblock {\em {The Emergent Multiverse: Quantum Theory According to the
  Everett Interpretation}}.
\newblock Oxford University Press, 2012.

\bibitem{Wallace:2020awr}
D.~Wallace.
\newblock ``{Against Wavefunction Realism}''.
\newblock In {\em {Current Controversies in Philosophy of Science}}, pages
  63--74. Routledge, 2020.

\bibitem{Weinberg:1996tqtfii}
S.~Weinberg.
\newblock {\em {The Quantum Theory of Fields}}, Volume~2.
\newblock Cambridge University Press, 1996.

\end{thebibliography}

\end{document}